\documentclass[aps,pre,twocolumn,groupedaddress,floatfix,10pt]{revtex4-2}

\usepackage{amsmath}
\usepackage{graphicx}
\usepackage{booktabs}
\usepackage[usenames,dvipsnames]{color}
\usepackage{tabularx}

\usepackage{xcolor}

\usepackage[cp1252]{inputenc}

\begin{document}


\title{Inertia-Driven Information Flow and Symmetry Breaking in a Nonequilibrium Two-Bead System}


\author{Jetin E. Thomas}
\email[]{jetinthomas@gmail.com}
\affiliation{Department of Physical Sciences, Indian Institute of Science Education and Research Mohali, \\ Knowledge city, Sector 81, Manauli, PO, Sahibzada Ajit Singh Nagar, Punjab 140306, India}
\author{Ramandeep S. Johal}
\email[]{rsjohal@iisermohali.ac.in}
\affiliation{Department of Physical Sciences, Indian Institute of Science Education and Research Mohali, \\ Knowledge city, Sector 81, Manauli, PO, Sahibzada Ajit Singh Nagar, Punjab 140306, India}


\date{\today}

\begin{abstract}
We investigate information-flow generation in a nonequilibrium
two-bead system coupled to two heat baths. We show that the
system acts as an information-flow generator in both overdamped and
underdamped regimes, with the underdamped dynamics revealing a divergence
of the scaled information flow along specific paths in the thermal
asymmetry--inertia parameter space that is hidden in the overdamped limit. This information generation suggests a possible route toward information engines and demon-like mechanisms in nanomachines. A symmetry-perturbation analysis of the response landscape of information flow reveals a geometric structure reminiscent of a Ginzburg--Landau framework: the symmetric reference state can correspond to a minimum or maximum depending on the perturbation direction, while the flat overdamped landscape develops a finite curvature under inertia. Mass asymmetry shifts the resulting maxima, and Hessian eigenvalue and eigenvector analysis reveals level touching of principal modes and bimodality along a constant-diffusion path. These results establish a minimal framework for understanding how inertia and microscopic heterogeneity shape information landscapes in nonequilibrium systems, with potential extensions to more complex heterogeneous networks.
\end{abstract}


\maketitle

\section{\label{Introduction} Introduction}
The concept of a Maxwell's demon posed a fundamental
question on the use of information  to exploit thermal
fluctuations and apparently circumvent the second law of thermodynamics.
The Szilard engine \cite{Szilard1929} subsequently established a quantitative connection
between information and thermodynamic entropy, while Landauer's principle clarified
the thermodynamic cost associated with irreversible information processing
\cite{Landauer1961,Bennett1982}. These developments established
the physical nature of information and laid the foundation for modern
information thermodynamics \cite{Maruyama2009,Parrondo2015}.

A particularly useful formulation arises for  bipartite stochastic systems
that can be partitioned into two interacting subsystems. Although the total system
obeys the conventional second law, the entropy balance of an individual
subsystem contains an additional contribution associated with the flow of
mutual information between the subsystems. Horowitz and Esposito \cite{HorowitzEsposito2014} established
a general stochastic-thermodynamic framework for this continuous information
flow, showing that information generated in one subsystem can modify the
thermodynamic balance of another and can enable processes that would
otherwise be forbidden by the local second law. Related bipartite formulations for autonomous Maxwell-demon
mechanisms, sensory systems, and information engines
\cite{HartichBaratoSeifert2014,BaratoSeifert2014} have shown that the information flow
provides a bridge between microscopic correlations, entropy production, and
energy transfer in coupled nonequilibrium systems.

This perspective is relevant for information engines in which correlations generated by
nonequilibrium fluctuations can be harnessed for work extraction. Recent work has further shown that heat engines and
information engines are not fundamentally separate classes of devices:
bipartite heat engines can operate through internal information transfer,
while information engines can benefit energetically from access to distinct
nonequilibrium reservoirs \cite{Leighton2024}. Recent studies have shown that
information-thermodynamic signatures and Maxwell-demon-like operation can
arise in molecular motors, such as kinesin under nonequilibrium
fluctuations \cite{LeightonSivak2025,DuBuisson2025}. These developments raise
a natural question: Rather than treating information flow only as a
thermodynamic correction or a quantity to be measured after the dynamics are
specified, can one understand the physical parameters that generate and
enhance information flow in the first place?

In this paper, we address this question using a minimal two beads and springs system
in simultaneous contact with two heat reservoirs. The model is sufficiently simple to permit an
analytic treatment of the steady-state covariance matrix, information
flow, heat currents, and entropy production, while retaining the essential
ingredients of interacting nonequilibrium dynamics. We show that the system
can act as an information-flow generator in both overdamped and underdamped
regimes. In the underdamped dynamics, inertia reveals regions of enhanced
information flow along specific paths in the thermal-asymmetry--inertia
parameter space that are hidden in the overdamped limit. The generated
information is not actively harnessed in the present autonomous setup, but
the results suggest a natural route toward information-engine or
Maxwell-demon-like operation through state-dependent coupling or jump
protocols \cite{DuBuisson2025}.


Information flow in coupled nonequilibrium systems has been studied in
relation to thermodynamic efficiency \cite{AllahverdyanJanzingMahler2009},
effective thermodynamics of underdamped particles
\cite{HerpichShayanfardEsposito2020}, energetic coordination in molecular
motors \cite{TakakiMugnaiThirumalai2022}, and information--work conversion
in bipartite heat engines \cite{LeightonEhrichSivak2024}. In contrast, we
investigate how information flow itself is organized by microscopic
symmetry breaking and inertia. We develop a multidimensional
information-flow landscape and show that its extrema, curvature,
bimodality, and principal modes are qualitatively reshaped by inertial and
mass asymmetries, providing a geometric framework for characterizing
information-generating states in nonequilibrium systems.

The proposed framework also provides a starting point for exploring whether response to 
information-flow landscapes can play a role analogous to free-energy
landscapes in equilibrium theories. The perturbative results obtained here,
including flat directions, shifted extrema, bimodality, and changes in
principal curvatures, suggest a Ginzburg--Landau-like description of
nonequilibrium response landscapes of information flow. While $\dot{\mathcal I}_{X}$ is not a free energy, extending this geometric framework to more complex heterogeneous
systems may provide a systematic way to identify preferred nonequilibrium
states and their preference to information flow generation.

The paper is organized as follows. In Sec.~\ref{sec:model}, we introduce the
two-bead model and its overdamped and underdamped dynamics. In
Sec.~\ref{sec:information_flow}, we derive the information flow and entropy
production and analyze their thermodynamic constraints. Section~\ref{subsec:OD}
presents the perturbative analysis of the overdamped information-flow
landscape, while Sec.~\ref{subsec:UD} extends it to the underdamped regime and
examines the effects of inertia and mass asymmetry. Finally, we discuss the
implications and possible extensions of these results in
Sec.~\ref{sec:Discussion}. Together, these analyses establish a minimal
framework for relating information generation, thermodynamic constraints,
symmetry breaking, and the geometry of nonequilibrium responses.

\section{\label{Model} Model}\label{sec:model}
\subsection{Two-bead nonequilibrium model}
Our system consists of two beads of masses $m_1$ ($m_2$), each directly coupled to a heat bath at temperature $T_1$ ($T_2$), with spring constant $k_{1}$  ($k_{2}$) and frictional drag coefficient $\gamma_1$ ($\gamma_2$). The two beads are harmonically coupled to each other with coupling constant $\kappa$ (see Fig. 1). 
\begin{figure}
\includegraphics[scale=0.4]{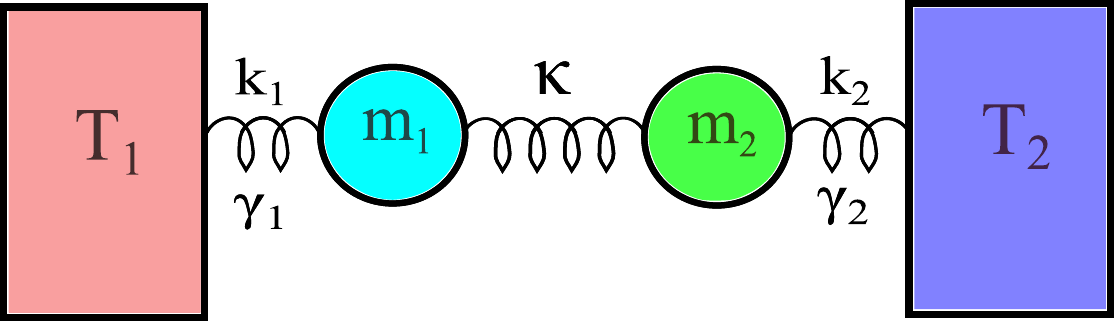}
\caption{Schematic of the two-beads and springs system in contact with 
two heat reservoirs at different temperatures.}
\label{twobeadssetup}
\end{figure}

 The Langevin equations \cite{saito2007fluctuation, kundu2011large, fogedby2012heat} for this system are 
\begin{align}
\frac{dx_{1}}{dt} = & \;  v_{1}, \label{pos1_Langevin} \\
\frac{dv_{1}}{dt} = & -\frac{k_{1}x_{1}}{m_{1}}
+\frac{\kappa(x_{2}-x_{1})}{m_{1}} 
-\frac{\gamma_{1}v_{1}}{m_{1}}+\xi_{1}(t), \label{vel1_Langevin}\\
\frac{dx_{2}}{dt} = & \;  v_{2}, \label{pos2_Langevin} \\
\frac{dv_{2}}{dt} = & 
-\frac{k_{2}x_{2}}{m_{2}}
-\frac{\kappa(x_{2}-x_{1})}{m_{2}} 
-\frac{\gamma_{2}v_{2}}{m_{2}}+\xi_{2}(t),
 \label{vel2_Langevin}
\end{align}
where $x_{1} (x_{2})$ and $v_{1} (v_{2})$ 
respectively are the displacement from the equilibrium position
and the velocity 
of the bead with mass $m_1$ ($m_2$). The white noise due to a bath satisfies: $\langle \xi_{\alpha}(t) \rangle = 0$ and $\langle \xi_{\alpha}(t)\xi_{\alpha}(t') \rangle = 2{\gamma_{\alpha}T_{\alpha}}\delta(t-t')/m_{\alpha}^2$, with $\alpha=1,2$.

\subsection{Overdamped limit}
In many soft-matter and biological systems, the viscous damping exerted by the surrounding medium dominates over inertial effects, causing the momentum relaxation time, $\tau_{m}=m/\gamma_{1,2}$,  to become much shorter than the configurational relaxation time, $\tau_{r}=\gamma_{1,2}/\kappa$   \cite{gardiner2009stochastic,risken1996fokker,seifert2012stochastic,doi1986theory}. When $\tau_{m}\ll\tau_{r}$, the velocities rapidly equilibrate with the surrounding fluid and can be adiabatically eliminated, yielding an overdamped description of the dynamics \cite{gardiner2009stochastic,risken1996fokker}. 
Thus, the overdamped limit provides a natural reference point for investigating how finite inertia modifies information transport and thermodynamic performance in nonequilibrium systems.

Accordingly, in the limit $m_{1},m_{2}\rightarrow0$, eqs.~(\ref{pos1_Langevin})--(\ref{vel2_Langevin}) reduce to
\begin{align}
\gamma_{1}\frac{dx_{1}}{dt}
&=
-k_{1}x_{1}
+\kappa(x_{2}-x_{1})
+\sqrt{2\gamma_{1}T_{1}}\xi_{1}(t), \label{pos1_OD_langevin}\\
\gamma_{2}\frac{dx_{2}}{dt}
&=
-k_{2}x_{2}
-\kappa(x_{2}-x_{1})
+\sqrt{2\gamma_{2}T_{2}}\xi_{2}(t), \label{pos2_OD_langevin}
\end{align}
where 
$\langle\xi_{\alpha}(t)\rangle=0$, and 
$\langle\xi_{\alpha}(t)\xi_{\beta}(t')\rangle
=\delta_{\alpha\beta}\delta(t-t')$, 
with $\alpha,\beta=1,2$. The overdamped dynamics is governed solely by the configurational degrees of freedom, with the interplay between the spring constants, friction coefficients, and thermal driving determining the nonequilibrium steady state. Throughout this work, we investigate both the overdamped and underdamped regimes to isolate the influence of inertia on information flow generation, entropy production, and the landscape of the response functions. As will be clear later, the ratio of the two relaxation times emerges naturally as one of the fundamental control parameters governing the crossover between overdamped and underdamped information flow.

\section{Information Flow and the Local Second Law}\label{sec:information_flow}

The rate of total entropy production for the coupled two-bead system obeys the conventional second law of thermodynamics and remains non-negative in the nonequilibrium steady state \cite{Seifert2012, Esposito2010}. However, when the system is partitioned into two interacting subsystems, the local entropy balance of each bead is modified by the continuous exchange of information between them. Consequently, the entropy production associated with an individual bead is no longer determined solely by the heat exchanged with its respective thermal reservoir but must also account for the information transferred between the coupled degrees of freedom \cite{HorowitzEsposito2014, HartichBaratoSeifert2014}. In this section, we derive the information flow for the overdamped two-bead system and demonstrate its role in
restoring the local second-law inequalities for each subsystem.

\subsection{Local second law for interacting subsystems}

Following Ref. \cite{HorowitzEsposito2014},
the local rates of entropy production satisfy
\begin{align}
\dot{\Sigma}_{X_{1}} &= \dot{S}_{X_{1}}-\frac{\dot{Q}_{X_{1}}}{T_1}-\dot{I}_{X_{1}},\\
\dot{\Sigma}_{X_{2}} &= \dot{S}_{X_{2}}-\frac{\dot{Q}_{X_{2}}}{T_2}-\dot{I}_{X_{2}},
\end{align}
where $\dot I_{X_2}$ ($\dot I_{X_1}$) denotes the information flow from $X_2$ to $X_1$ ($X_1$ to $X_2$), and $\dot Q_{X_2}$ ($\dot Q_{X_1}$) denotes the corresponding heat flux. Here, $X_1$ and $X_2$ are coupled to heat baths at temperatures $T_1$ and $T_2$, respectively, and $k_B=1$. The local rates of entropy production and Shannon entropy for each bead are denoted by $\dot{\Sigma}_{X_{i}}$ and $\dot{S}_{X_{i}}$, respectively. In the nonequilibrium steady state, the system entropies become time independent, i.e. $\dot{S}_{X_{i}} =0$, while $\dot{\Sigma}_{X_{i}}>0$ and $\dot I_{X_{1}}+\dot I_{X_{2}} =0$ \cite{HorowitzEsposito2014}, thus reducing the local second-law inequalities to
\begin{align}
\dot{\Sigma}_{X_{1}} &= -\frac{\dot{Q}_{X_{1}}}{T_1}+\dot{I}_{X_{2}}\ge0, \label{local_EP_ineq_1}\\
\dot{\Sigma}_{X_{2}} &= -\frac{\dot{Q}_{X_{2}}}{T_2}-\dot{I}_{X_{2}}\ge0. \label{local_EP_ineq_2}
\end{align}
The explicit expressions for the information flow obtained in eqs. (\ref{IdotY_OD}) and (\ref{IdotY_UD}) exactly satisfies the local entropy balance relations above. Thus, information flow accounts for the apparent local Clausius inequality, while it cancels upon summing the two subsystem balances and the conventional second law for the composite system is recovered,
\begin{equation}
\dot\Sigma
=
-\frac{\dot{Q}_{X_{1}}}{T_{1}}
-\frac{\dot{Q}_{X_{2}}}{T_{2}}
\ge0,
\label{sigdot}
\end{equation}
%
\subsection{Steady-state information flow}
The steady-state heat currents exchanged with the thermal reservoirs satisfy energy conservation, and
with the explicit expression from \cite{Thomas2026Kinetic} and taking the overdamped limit ($m_{1},m_{2}\rightarrow 0$)
\begin{equation}
\dot{Q}_{X_{1}}
=
-\dot{Q}_{X_{2}}
=
\frac{
(T_{1}-T_{2})\kappa^{2}
}{
\gamma_{2}(k_{1}+\kappa)
+
\gamma_{1}(k_{2}+\kappa)
}.
\label{Qdot_OD}
\end{equation}

The thermal conductivity between the heat baths ($\lambda_{bath}$) is given as,

\begin{equation}
\lambda_{bath}
=
\frac{\dot{Q}_{X_{1}}}{(T_{1}-T_{2})}
=
\frac{
\kappa^{2}
}{
\gamma_{2}(k_{1}+\kappa)
+
\gamma_{1}(k_{2}+\kappa)
}.
\label{lambda_bath_OD}
\end{equation}

The information flow from bead $X_{2}$ to $X_{1}$ is given by

\begin{equation}
\dot I_{X_{2}}=\frac{
\kappa^{2}(T_{1}-T_{2})
\left(
\gamma_2(k_{1}+\kappa)T_2
+
\gamma_1(k_{2}+\kappa)T_1
\right)
}
{\mathcal{D}en},
\label{IdotY_OD}
\end{equation}
where 
\begin{equation}
\begin{aligned}
\mathcal{D}en\equiv
&(\gamma_{2}k_{1}+\gamma_{1}k_{2})^{2}T_{1}T_{2}
+2\kappa(\gamma_{1}+\gamma_{2})
(\gamma_{2}k_{1}+\gamma_{1}k_{2})
T_{1}T_{2}\\
&+
\kappa^2(\gamma_{2}T_{1}+\gamma_{1}T_{2})
(\gamma_{1}T_{1}+\gamma_{2}T_{2}).
\end{aligned}
\label{D_OD}
\end{equation}
The detailed derivation of eqs.~(\ref{IdotY_OD}-\ref{D_OD}) is presented in Appendix~B.
The local entropy production rates are obtained 
from eqs. (9) and (10). 

The explicit forms of $\dot{\Sigma}_{X_{1}}$ and $\dot{\Sigma}_{X_{2}}$ are manifestly non-negative for all physically admissible parameter values, confirming that the information flow precisely accounts for the local entropy balance while being canceled from the total entropy production.  Having established the exact thermodynamic relations governing information flow in the overdamped system, we next investigate how the information-generating state evolves under perturbations of the microscopic system parameters and thermal drive.

\subsection{Near-equilibrium expansion and entropy-production scaling}
\label{sec:near_equilibrium}

It is useful to examine the information flow and entropy production in the
vicinity of the equilibrium reference state.
We parametrize the bath temperatures as 
\begin{equation}
T_{1,2}=T_0\left(1\pm\frac{\delta}{2}\right),
\label{t12}
\end{equation}
where $T_0 = (T_1+T_2)/{2}$ and $-2\leq \delta \leq 2$.
$\delta=0$ corresponds to the equilibrium situation, at which
both the heat current and information flow vanish.

The information flow and the heat contribution to the local entropy balance
can be expanded about $\delta=0$, as 
\begin{align}
\dot I_{X_2}
=
\delta\left(\frac{\partial \dot{I}_{X_{2}}}{\partial\delta}\Big{|}_{\delta=0} \right. & \left. + \frac{1}{2!} \frac{\partial^{2} \dot{I}_{X_{2}}}{\partial\delta^{2}}\Big{|}_{\delta=0}\delta \right.\nonumber\\
\left. \right.&\left.+ \frac{1}{3!}\frac{\partial^{3} \dot{I}_{X_{2}}}{\partial\delta^{2}}\Big{|}_{\delta=0}\delta^2
+\mathcal{O}(\delta^3)\right),
\label{eq:I_delta_expansion}
\\
\frac{\dot Q_{X_{1}}}{T_1}=\frac{\lambda_{bath}\delta}{(1+\delta/2)}
=&\lambda_{bath}\delta\left(1-\frac{\delta}{2}+\frac{\delta^2}{4}+\mathcal{O}(\delta^3)\right),
\label{eq:Q1_delta_expansion}
\\
-\frac{\dot Q_{X_{2}}}{T_2}=\frac{\lambda_{bath}\delta}{(1-\delta/2)}
=&\lambda_{bath}\delta\left(1+\frac{\delta}{2}+\frac{\delta^2}{4}+\mathcal{O}(\delta^3)\right).
\end{align}
The absence of a constant term follows from the vanishing of the heat current and information flow at equilibrium as in eqs. (\ref{Qdot_OD}) and (\ref{IdotY_OD}) respectively. Moreover, the leading terms in the two quantities are constrained by the local thermodynamic
balance for the entropy production of subsystem $X_{1,2}$ in eq. (\ref{local_EP_ineq_1}-\ref{local_EP_ineq_2}). We can exactly solve the subsystem entropy production from eqs.~(\ref{local_EP_ineq_1}-\ref{local_EP_ineq_2}) using eqs.~(\ref{Qdot_OD}) and (\ref{IdotY_OD}) or independently from eqs.~(\ref{Sigma_X1}-\ref{Sigma_X2}) and observe they go as $(T_{1}-T_{2})^{2}$. Therefore, entropy production must
vanish at equilibrium and begins at second order for series expanded in terms of the thermodynamic driving ($\delta$), hence, the linear term must cancel, giving
\begin{equation}
\frac{\partial \dot{I}_{X_{2}}}{\partial \delta}\Big{|}_{\delta=0}=\lambda_{bath}.
\label{eq:Idot_lbath_rel}
\end{equation}
Consequently,
\begin{align}
\dot\Sigma_{X_{1,2}} = & \delta^2\left(\left(\frac{\lambda_{bath}}{2}\pm\frac{\partial^{2}\dot{I}_{X_{2}}}{2!\partial\delta^{2}}\Big{|}_{\delta=0}\right) \nonumber\right.\\
\left.\right.&\left.\pm\left(\frac{\partial^{3}\dot{I}_{X_{2}}}{3!\partial\delta^{3}}\Big{|}_{\delta=0}-\frac{\lambda_{bath}}{4}\right)\delta
+\mathcal{O}(\delta^2)\right),
\label{eq:sigma_X2_expansion}
\end{align}

Thus, $\dot\Sigma_{X_{1,2}}$ vanishes quadratically with the thermal driving near equilibrium, so that limit of $\dot\Sigma_{X_{1,2}}/\delta^2$ remains finite as $\delta\to0$. Similarly, $\dot{I}_{X_{1,2}}$ vanishes linearly with $\delta$ so that limit of $\dot{I}_{X_{1,2}}/\delta$ remains finite as $\delta\to0$ from eq.~(\ref{eq:I_delta_expansion}).

Importantly, this structure does not depend on whether the dynamics is
overdamped or underdamped. Inertia changes the coefficients in the
expansions through the covariance matrix and the additional mass-dependent
timescales, but the equilibrium conditions
$\dot I_{X_i}=0$ and $\dot\Sigma_{X_i}=0$ and the resulting cancellation of
the linear entropy-production term remain unchanged. 

\section{Information flow generation}
At thermal equilibrium ($T_{1}=T_{2}$), the steady-state probability currents vanish, and there is no directed transfer of information between the two beads, i.e., $\dot{I}_{X_{2}}=0$. A non-zero temperature difference breaks detailed balance and establishes a persistent heat current through the harmonic coupling. As a consequence, the stochastic dynamics of the two beads become increasingly correlated, leading to a continuous increase in the information transferred from one subsystem to the other. 

\begin{figure}[t] 
\centering 
\includegraphics[width=0.95\columnwidth]{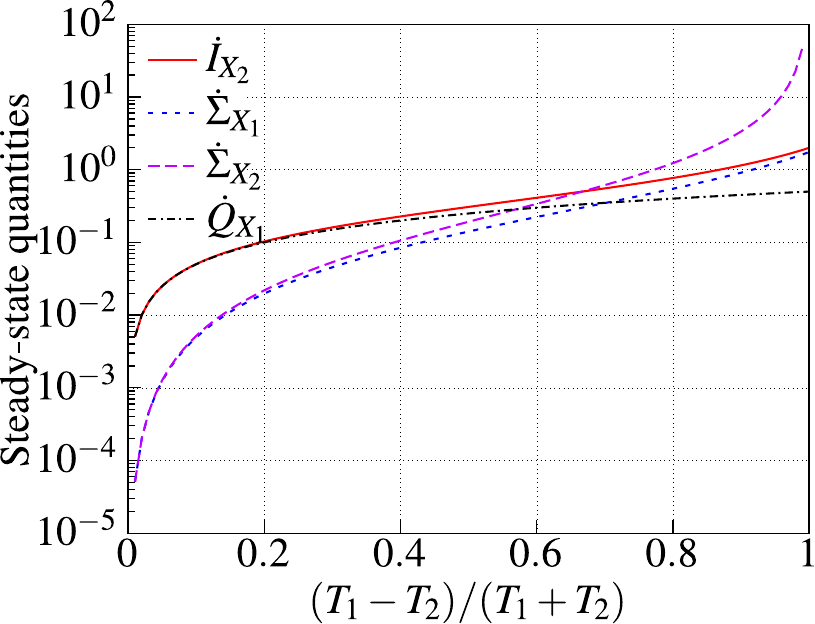} 
\caption{Steady-state information flow $\dot{I}_{X_{2}}$, subsystem entropy production rates $\dot{\Sigma}_{X_{1}}$ and $\dot{\Sigma}_{X_{2}}$, and heat current $\dot{Q}_{X_{1}}$ as functions of the normalized temperature difference ($\xi$) for the symmetric system (all parameters set to unity) as informational flow only depends on $\xi$ and not on the mean of the temperatures of the two heat baths ($T_{0}$).} 
\label{fig:InformationGeneration} 
\end{figure}

Figure~\ref{fig:InformationGeneration} shows the steady-state information flow $\dot{I}_{X_{2}}$ together with the subsystem entropy production rates and heat current as functions of the thermal driving. 

As expected from eq.~(\ref{IdotY_OD}), the information flow increases monotonically with the temperature difference, demonstrating that stronger nonequilibrium forcing enhances the exchange of information between the coupled degrees of freedom. The simultaneous increase of the heat current and subsystem entropy production rates confirms that stronger thermal driving not only transports more energy through the system but also establishes stronger dynamical correlations between the two interacting beads.

\subsection{Information flow as a measure of dynamical coordination}
The increase in information flow is accompanied by corresponding increases in the local entropy production rates, $\dot{\Sigma}_{X_{1}}$ and $\dot{\Sigma}_{X_{2}}$, as well as the heat current exchanged with the reservoirs. This behavior follows directly from the local entropy balance relations (eqs \ref{local_EP_ineq_1} and \ref{local_EP_ineq_2}), which show that the information flow provides the additional contribution required for each subsystem to satisfy the second law individually. Consequently, stronger information transfer reflects a greater degree of nonequilibrium coordination between the two beads, while the accompanying increase in the subsystem entropy production quantifies the additional irreversible dissipation required to sustain this coordinated dynamics.

These observations demonstrate that the information flow is not just meant for bookkeeping to restore the local second law. Rather, it provides a quantitative measure of the dynamical coordination established between the two beads through the combined action of thermal driving and harmonic coupling. As the temperature difference increases, the system generates progressively larger amounts of information, indicating that nonequilibrium energy supplied by the thermal reservoirs is continuously converted into directed information transfer between the interacting subsystems. In this sense, the two-bead system constitutes a minimal autonomous information generator, sharing key characteristics with autonomous information-processing systems, biochemical sensory networks, molecular motors, and Maxwell-demon-type information engines, where persistent energy dissipation sustains continuous information exchange and enables energy transduction. \cite{HorowitzEsposito2014,HartichBaratoSeifert2014,Barato2014,Mandal2012,Parrondo2015}. Unlike feedback-controlled information engines, no external measurement or control protocol is required here; the information is generated autonomously through the interplay of thermal driving and harmonic coupling.

\section{A Perturbative Analysis}
While the temperature gradient determines the overall strength of information flow generation, the efficiency with which information is generated also depends on the microscopic properties of the coupled system, including the spring constants, friction coefficients, and masses. This naturally raises two fundamental questions: Which combinations of microscopic parameters maximize information flow generation? How robust are these information-generating states against symmetry-breaking perturbations? To address these questions, we perform a perturbative analysis of the response landscape of information flow by breaking the symmetry of the symmetric equilibrium configuration. We define the reference symmetric equilibrium configuration as the configuration with system parameters which are symmetrical i.e. the spring constant of side springs are equal ($k_{1}=k_{2}$), the frictional damping coefficient on the beads are equal ($\gamma_{1}=\gamma_{2}$) and the temperature difference between the two heat baths are equal ($T_{1}=T_{2}$). For underdamped system, the symmetric reference state referred there have the additional constraint that the masses of two beads are equal ($m_{1}=m_{2}$).


\subsection{Overdamped information landscape}
\label{subsec:OD}
The quantities $\dot{I}_{X_{i}}$ and $\dot{\Sigma}_{X_{i}}$ contain overall multiplicative factors proportional to the nonequilibrium driving, namely $(T_{1}-T_{2})$ and $(T_{1}-T_{2})^2$, respectively. These factors determine the absolute magnitude of information flow generation and dissipation, but do not affect their dependence on the intrinsic microscopic properties of the system. To isolate the geometry of the response landscape of information flow from the overall strength of the thermal driving, we introduce the following scaled dimensionless quantities. 
\begin{equation}
\left.
\begin{aligned}
\bar{I}_{X_{2}}
&=
\frac{\dot I_{X_{2}}}{T_{1}-T_{2}},\\
\bar{\Sigma}_{X_{i}}
&=
\frac{\dot{\Sigma}_{X_{i}}}{(T_{1}-T_{2})^{2}}.
\end{aligned}
\right\} \quad\text{Scaled steady-state quantities.}
\label{ScaledQuantities}
\end{equation}
%
The above scaled quantities depend on the six
independent microscopic parameters
$(k_{i}'s,\gamma_{i}'s,T_{i}'s)$.
Next, to investigate how information flow generation responds to
symmetry-breaking perturbations, we introduce additional 
parameterizations:
\begin{align}
k_{1,2}&=\lambda\kappa\left(1\pm\frac{\alpha}{2}\right), 
\quad \gamma_{1,2}=\gamma_{0}\left(1\pm\frac{\epsilon}{2}\right), 
\label{PerturbationParametersOD}
\end{align}
analogous to eq.~(\ref{t12}). Here, $\lambda=k_{0}/\kappa$ measures the average stiffness, $k_{0} =(k_1+k_2)/2$,  of the
side springs relative to the central spring, while
$\alpha$ and $\epsilon$ quantify the relative asymmetry
in the respective parameters. 
Note that the symmetric reference configuration corresponds to
$\alpha=\epsilon=\delta=0$.
Substituting eqs.~(\ref{PerturbationParametersOD}) into 
eq.~(\ref{ScaledQuantities}), the dependence on the microscopic
parameters can be written as
\begin{align}
\bar I_{X_{2}}
&=
\frac{\kappa}{\gamma_{0}T_{0}}
\,\dot{\mathcal I}_{X_{2}}(\alpha,\epsilon,\delta;\lambda),
\label{ScI_OD}\\
\bar\Sigma_{X_{i}}
&=
\frac{\kappa}{\gamma_{0}T_{0}^2}
\,\dot{\mathcal S}_{X_{i}}(\alpha,\epsilon,\delta;\lambda),
\end{align}
where $\dot{\mathcal I}_{X_{2}}$, $\dot{\mathcal S}_{X_{1}}$, and $\dot{\mathcal S}_{X_{2}}$ are
dimensionless functions describing the response landscape of information flow.
The overall dependence on the coupling spring constant scales out nicely,
leaving the scaled quantities to depend only on the stiffness ratio
$\lambda$ and the symmetry-breaking parameters.
Consequently, the perturbation analysis presented below focuses on the
geometry of 
$\dot{\mathcal I}_{X_{2}}$, $\dot{\mathcal S}_{X_{1}}$, and $\dot{\mathcal S}_{X_{2}}$, whose local extrema
and curvature characterize the sensitivity of information flow generation
and subsystem entropy production to microscopic perturbations. Note, the scaled quantities in eqs.~(\ref{ScaledQuantities}) should be considered as response functions of the system's information flow and local entropy production on thermal driving away from equilibrium. Even though, for equilibrium configurations ($\delta=0$) they are not defined or couldn't be measured but their limit exists on approaching equilibrium, Exactly analogous to conductance emerging on applying tiny potential difference across ohmic resistors which couldn't be measured when the circuit is switched off.

\subsection{Perturbation along independent directions}
We begin by examining perturbations along the three independent coordinate directions defined by the symmetry-adapted parameters  $(\alpha,\delta,\epsilon)$. By varying a single parameter at a time while keeping the other two fixed at their null values, the local response can be resolved along each coordinate axis. For the cut along $\alpha$, keeping the other parameters $\delta$ and $\epsilon$ equal to 0. We have from eqs.~(\ref{eq:Idot_lbath_rel}-\ref{eq:sigma_X2_expansion}),
%
\begin{figure}[t]
    \centering
   \includegraphics[width=1.0\columnwidth,keepaspectratio]{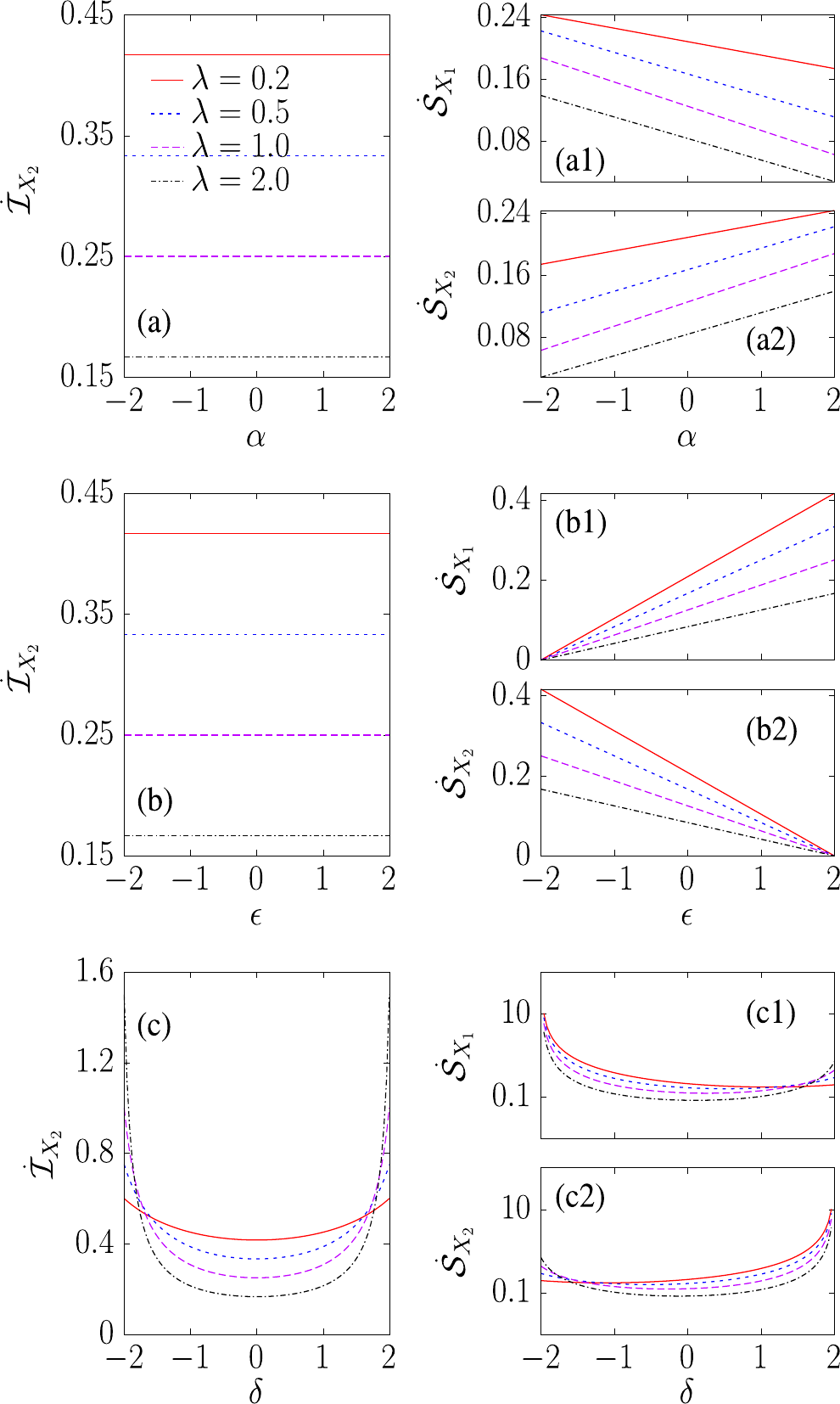}
    \caption{
    Perturbative response landscapes of information flow for the information flow $\dot{\mathcal{I}}_{X_{2}}$ (left column) and subsystem entropy productions $\dot{\mathcal{S}}_{X_{1}}$ and $\dot{\mathcal{S}}_{X_{2}}$ (right column). Each row corresponds to perturbations along the spring asymmetry ($\alpha$), friction asymmetry ($\epsilon$), and temperature asymmetry ($\delta$) directions, respectively, while the remaining perturbation parameters are held fixed at zero. Curves are shown for $\lambda=0.2$, $0.5$, $1$, and $2$. 
    }
    \label{fig:PerturbationLandscape}
\end{figure}

\begin{align}
\frac{\kappa}{\gamma_{0}}\dot{\mathcal{I}}(\mathbf{x})
\equiv
\left.
\frac{\partial \dot{I}_{X_2}}{\partial \delta}
\right|_{\mathbf{x}}
&=
\lambda_{\mathrm{bath}}(\mathbf{x})
=
\frac{\kappa}{2\gamma_{0}(1+\lambda)},\label{ScI_alpha_cut}
\\[4pt]
\frac{\kappa}{\gamma_{0}}\mathcal{S}_{X_{1,2}}(\mathbf{x})
&=
\lambda_{\mathrm{bath}}(\mathbf{x})
\frac{\left(2+(2\mp\alpha)\lambda\right)}
{4(1+\lambda)},\label{ScS12_alpha_cut}\\
&\hspace{2.5cm}\text{where}\;
\mathbf{x}=(\alpha,0,0).
\nonumber
\end{align}
For the cut along $\epsilon$, keeping the other parameters $\delta$ and $\alpha$ equal to 0. We have from eqs.~(\ref{eq:Idot_lbath_rel}-\ref{eq:sigma_X2_expansion}),

\begin{align}
\frac{\kappa}{\gamma_{0}}\dot{\mathcal{I}}(\mathbf{x})
\equiv
\left.
\frac{\partial \dot{I}_{X_2}}{\partial \delta}
\right|_{\mathbf{x}}
&=
\lambda_{\mathrm{bath}}(\mathbf{x})
=
\frac{\kappa}{2\gamma_{0}(1+\lambda)},\label{ScI_epsilon_cut}
\\[4pt]
\frac{\kappa}{\gamma_{0}}\mathcal{S}_{X_{1,2}}(\mathbf{x})
&=
\lambda_{\mathrm{bath}}(\mathbf{x})
\frac{\left(2\pm\epsilon\right)}
{4}, \label{ScS12_epsilon_cut}
\\[4pt]
&\hspace{2.5cm}\text{where}\;
\mathbf{x}=(0,0,\epsilon).
\nonumber
\end{align}

For the cut along $\delta$, keeping the other parameters $\alpha$ and $\epsilon$ equal to 0, i.e. for a symmetric configuration. We have from eqs.~(\ref{eq:I_delta_expansion},\ref{eq:Idot_lbath_rel}-\ref{eq:sigma_X2_expansion}),

\begin{align}
\frac{\kappa}{\gamma_{0}}\dot{\mathcal{I}}(\mathbf{x})
&=\lambda_{\mathrm{bath}}(\mathbf{x})\left(1+\frac{\lambda(2+\lambda)}{4(1+\lambda)^{2}}\delta^{2} + \mathcal{O}(\delta^3)\right),\label{ScI_delta_cut}
\\[4pt]
\frac{\kappa}{\gamma_{0}}\mathcal{S}_{X_{1,2}}(\mathbf{x})
&=
\lambda_{\mathrm{bath}}(\mathbf{x})
\left(\frac{1}
{2}\mp\frac{\delta}{4(1+\lambda)^2}+\frac{\delta^2}{8}+\mathcal{O}(\delta^3)\right),\label{ScS12_delta_cut}
\\[4pt]
&\text{where}\;
\lambda_{\mathrm{bath}}(\mathbf{x})=\frac{\kappa}{2\gamma_{0}(1+\lambda)}\; \text{and}\; \mathbf{x}=(0,\delta,0).\nonumber
\end{align}

Figure \ref{fig:PerturbationLandscape} shows that the scaled information flow exhibits markedly different sensitivities depending on the perturbation direction. Perturbations in the spring and friction asymmetries produce only flat variations of scaled information flow over the parameter range considered (eqs.~(\ref{ScI_alpha_cut}) and (\ref{ScI_epsilon_cut})), whereas perturbations in the temperature asymmetry generate a pronounced nonlinear response with positive curvature (positive coefficient at $\mathcal{O}(\delta^{2})$ in eq.~(\ref{ScI_delta_cut})) and a minimum (zero coefficient for $\mathcal{O}(\delta)$ in eq.~(\ref{ScI_delta_cut})) for the parameters at the symmetric configuration at equilibrium. The subsystem entropy production rates are considerably more sensitive to symmetry breaking. Along the spring and friction directions they vary linearly (eq.~(\ref{ScS12_alpha_cut}) and eq.~(\ref{ScS12_epsilon_cut})), indicating a redistribution of dissipation between the two subsystems, while thermal perturbations lead to highly nonlinear variations spanning several orders of magnitude (eq.~(\ref{ScS12_delta_cut})). The perturbation slices of Fig.~\ref{fig:PerturbationLandscape} characterize the response along the coordinate axes only. To determine the intrinsic local geometry of the response landscape of information flow, we now analyze the Hessian.

\subsection{Hessian analysis of the response landscape}
\begin{figure}[t]
\centering
\includegraphics[width=\linewidth]{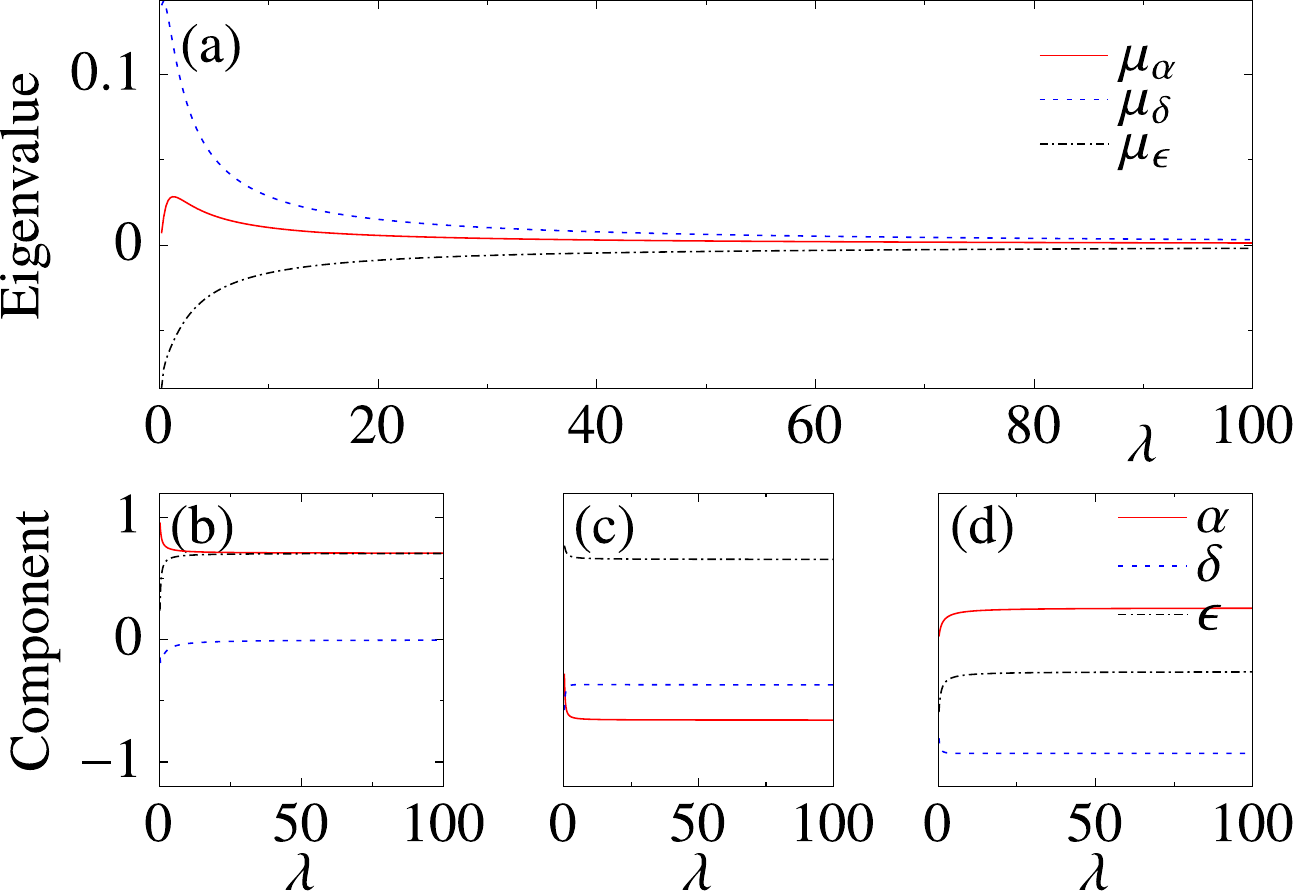}
\caption{
Principal curvature analysis of the overdamped response landscape as a function of the dimensionless coupling parameter $\lambda$. 
(a) Eigenvalues of the Hessian evaluated at the symmetric steady state
$(\alpha,\delta,\epsilon)=(0,0,0)$.
The coexistence of one negative and two positive eigenvalues demonstrates that the symmetric state is a saddle point of the three-dimensional perturbation landscape. 
(b--d) Components of the normalized eigenvectors corresponding to the weakly stable, unstable, and strongly stable modes, respectively, as explained in Sec. V-D. 
}
\label{fig:HessianModes}
\end{figure}

While the perturbation slices shown in Fig.~\ref{fig:PerturbationLandscape}
characterize the response along the coordinate axes, they do not reveal the
intrinsic directions along which the response landscape varies.
We examine the Hessian matrix of the scaled information flow
by characterizing the local geometry of the response landscape around the symmetric reference state $(\alpha,\delta,\epsilon)=(0,0,0)$, 
\begin{equation}
H_{ij}
=
\left.
\frac{\partial^2\dot{\mathcal I}_{X_{2}}}
{\partial x_i\partial x_j}
\right|_{\mathbf{x}=\mathbf{0}},
\qquad
\mathbf{x}=(\alpha,\delta,\epsilon)^{\mathrm T}.
\end{equation}
Expanding the information flow about the symmetric state gives
\begin{equation}
\dot{\mathcal I}_{X_{2}}(\mathbf{x})
=
\dot{\mathcal I}_{X_{2}}(\mathbf{0})
+\frac12
\mathbf{x}^{\mathrm T}
H
\mathbf{x}
+\mathcal{O}(|\mathbf{x}|^3),
\end{equation}
where the eigenvalues of $H$ determine the principal curvatures of the
response landscape and the corresponding eigenvectors define its natural
perturbation directions. Direct differentiation shows that
\[
\nabla\dot{\mathcal I}_{X_{2}}\big|_{(\alpha,\delta,\epsilon)=(0,0,0)}
=\mathbf0,
\]
so that the linear term vanishes and the quadratic term provides the leading
variation around the symmetric state.
Figure~\ref{fig:HessianModes}(a) shows the evolution of the Hessian eigenvalues as the dimensionless coupling parameter $\lambda$ is varied. Beyond $\lambda \approx 20$, both the eigenvalues and eigenvectors vary only weakly with increasing coupling, indicating that the local geometry of the response landscape approaches an asymptotic strong-coupling regime. Figure~\ref{fig:HessianModes}(a) shows the evolution of the Hessian eigenvalues as $\lambda$ is varied. As the coupling strength increases, the magnitudes of all three eigenvalues decrease approximately as $1/\lambda$, implying that the quadratic curvature of the response landscape vanishes asymptotically in the strong-coupling limit. Consequently, the information flow becomes progressively less sensitive to infinitesimal symmetry-breaking perturbations. The presence of one negative eigenvalue indicates that the symmetric state is locally unstable with respect to one collective perturbation direction, whereas the two positive eigenvalues correspond to restoring directions. Consequently, the symmetric configuration is a saddle point rather than a local minimum of the response landscape. The limit $\lambda\rightarrow0$ is singular,  because the side springs vanish and the center-of-mass coordinate becomes unconfined, preventing the existence of a normalizable stationary distribution. This does not imply the absence of dynamical correlations or information exchange between the coupled degrees of freedom. Rather, it indicates that the present steady-state response landscape is no longer the appropriate framework. Such situations naturally arise in molecular transport models, where the motor and the cargo remain elastically coupled while their overall motion is unconfined, and the relevant dynamics is governed by the relative coordinate and the nonequilibrium trajectory statistics, rather than by a globally stationary distribution \cite{Korn2009,Erickson2011,Goychuk2014}. 

\subsection{Local stability and physical interpretation}
The Hessian analysis provides considerably more information than the one-dimensional perturbation curves discussed previously. While the latter probe the curvature only along the coordinate directions, the eigenvectors identify the principal directions in parameter space along which the information flow varies independently.

The corresponding normalized eigenvectors are shown in Figs.~\ref{fig:HessianModes}(b)--(d). Since an eigenvector and its negative represent the same physical mode, the overall sign of each eigenvector is chosen to maintain continuity with respect to $\lambda$. Furthermore, the eigenvectors are tracked by maximizing their overlap with those obtained at the previous value of $\lambda$, ensuring a consistent identification of the principal perturbation modes throughout the parameter sweep.
The eigenvector associated with the smallest positive eigenvalue is dominated by correlated perturbations of the spring and friction asymmetries, with only a negligible contribution from the temperature asymmetry. This defines an almost weakly stable direction along which the information flow changes only weakly. In contrast, the eigenvector corresponding to the negative eigenvalue is approximately the orthogonal combination of the spring and friction perturbations, with a small admixture of the temperature asymmetry, thereby defining the unstable direction of the local response landscape. The remaining eigenvector, corresponding to the largest positive eigenvalue, is primarily aligned with the temperature asymmetry, indicating that thermal perturbations produce the strongest restoring curvature around the symmetric state.

These results provide a natural explanation for the perturbation landscapes
shown in Fig.~\ref{fig:PerturbationLandscape}. The flat dependence of the
information flow on the individual spring and friction perturbations arises
because neither coordinate axis coincides with a principal direction of the
Hessian. Instead, the natural perturbation coordinates are correlated
combinations of the microscopic asymmetry parameters. In contrast, the
temperature perturbation is closely aligned with one of the principal
directions and therefore exhibits a pronounced minimum at the symmetric
configuration.

Overall, 
these principal modes identify the
directions of strongest and weakest sensitivity of the non-dimensionalized information
flow around the symmetric reference state, thereby providing a natural
geometric basis for understanding how different microscopic asymmetries
cooperatively influence nonequilibrium information transfer. The analytical
framework developed here serves as a useful reference for interpreting the
more intricate perturbation landscape that emerges in the underdamped regime, as discussed below. 

\section{The underdamped regime}
\label{sec:UD}
The overdamped dynamics therefore provides a natural reference landscape against which the effects of inertia can be identified. We next extend the analysis to the underdamped regime and examine how the additional inertial and mass degrees of freedom modify information generation and its local landscape. Therefore, we solve the steady-state covariance matrix from the corresponding Lyapunov equation and substitute it into the general expression for the information flow (eq.~(\ref{eq:IdotCovariance})) in Appendix A. Unlike the overdamped case, here the inertia couples with the positional and velocity degrees of freedom, resulting in a considerably more intricate analytical expression. The steady-state information flow from subsystem $X_2$ to $X_1$ can nevertheless be written in a compact form:
\begin{equation}
\dot{I}_{X_{2}} = \frac{\gamma_{1}\gamma_{2}\kappa^{2}(T_{1}-T_{2})\mathcal{P}}{\mathcal{Q}},
\label{IdotY_UD}
\end{equation}
where both $\mathcal{P}$ and $\mathcal{Q}$
are multivariate polynomials of the system parameters 
$(m_{i},\gamma_{i},k_{i},\kappa,T_{i})$. Some notable features of the above representation 
are as follows. First, $\dot{I}_{X_{2}}$ is proportional to the temperature difference $T_{1}-T_{2}$, ensuring that the information flow vanishes at thermal equilibrium. Second, the information flow is proportional to $\kappa^2$, demonstrating that coupling between the subsystems is essential for information transfer. Finally, the dependence of $\mathcal{P}$ and $\mathcal{Q}$ on the masses $m_{1}$ and $m_{2}$ explicitly captures the inertial corrections absent in the overdamped limit.

\subsection{Universal scaling and inertial enhancement of information flow}
To elucidate the role of inertia on information transfer, we consider the symmetric reference state defined by 
$m_i=M_{0}, \gamma_i=\gamma_{0}, k_i=k_{0}$,
while allowing a nonzero thermal gradient ($\delta \neq 0$). Following the parametrization adopted in the overdamped analysis, we work with the dimensionless variables
\begin{equation}
\chi=\frac{M_{0}\kappa}{\gamma_{0}^2},\qquad
\lambda=\frac{k_{0}}{\kappa},\qquad
\frac{\delta}{2}=\frac{T_1-T_2}{T_1+T_2},
\end{equation}
where $\chi$ measures the relative importance of the inertia to viscous relaxation, and so on. For equal friction coefficients,
\begin{equation}
\frac{\delta}{2}=\frac{D_1-D_2}{D_1+D_2},
\end{equation}
with $D_i=k_BT_i/\gamma_{0}$. The normalized information flow assumes the universal scaling form
\begin{equation}
\begin{aligned}
&\frac{\gamma_{0}}{\kappa}\dot I_{X_{2}}
=\mathcal{F}(\delta,\chi,\lambda)
=\\
&\frac{\delta\left(4\chi+(4-\delta^2)(1+\lambda)\right)}
{8\chi^2+(4-\delta^2)(4(1+\lambda)\chi+\left(4+(4-\delta^2)\lambda(2+\lambda)\right))}.
\label{ScI_UF_DL}
\end{aligned}
\end{equation}
\begin{figure}[t]
\centering
\includegraphics[width=\columnwidth]{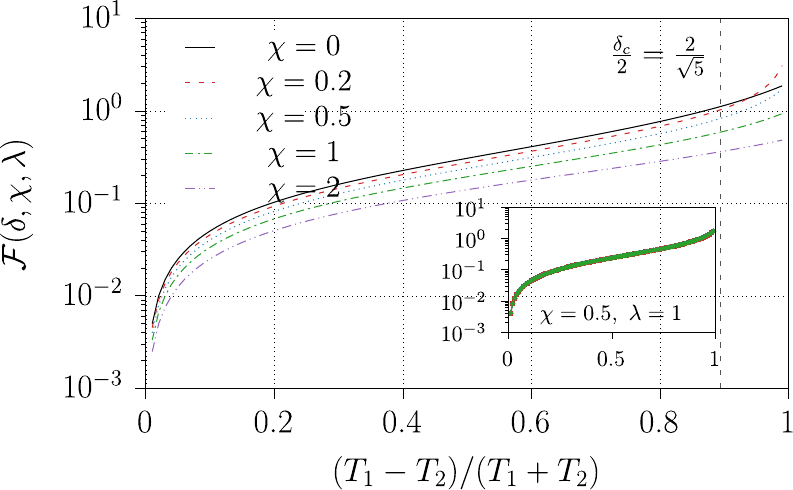}
\caption{
Universal scaling function $\mathcal{F}(\delta,\chi,\lambda)=\gamma\dot{I}_{X_{2}}/\kappa$ as a function of the nonequilibrium driving
($\delta$) for $\lambda=1$ and different values of the inertial parameter $\chi$. The inset shows the dependence of the scaling function on the non-equilibrium drive $\delta$ for fixed $\chi=0.5$ and $\lambda=1.0$. The vertical dashed line is to mark $\delta_c=4/\sqrt{5}$, the onset of inertia enhancement.
}
\label{fig:ScalingFunction}
\end{figure}

Figure~\ref{fig:ScalingFunction} summarizes the universal dependence of the information flow on the nonequilibrium driving and inertia. For weak driving ($\delta\ll1$), the overdamped limit maximizes the information flow. Above a critical driving, the dependence on $\chi$ becomes non-monotonic and a finite optimal inertia emerges. Irrespective of the value of $\chi$, the coupled two-bead system acts as an information generator, as evidenced by the monotonic increase of the information flow with the nonequilibrium driving.

\begin{figure}[htbp]
  \centering
  \includegraphics[width=\columnwidth]{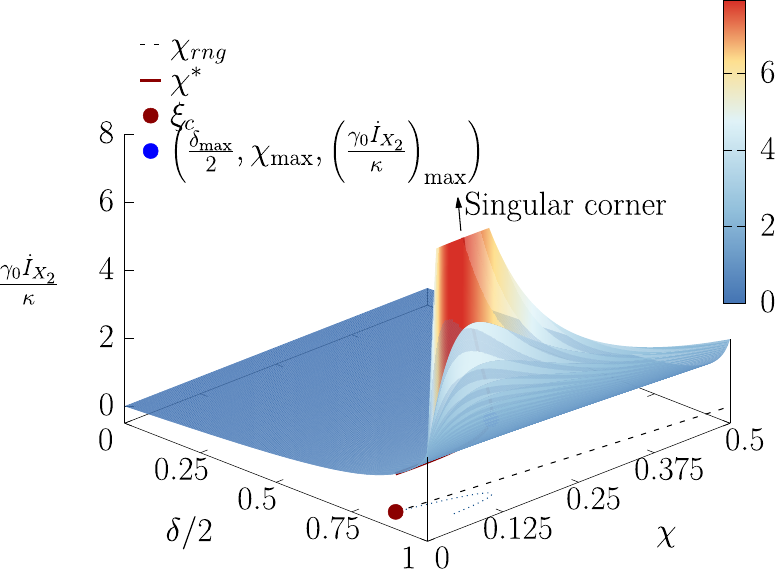}
  \caption{
Universal geometry of the information flow
$\mathcal{F}(\delta,\chi,\lambda)=\gamma_{0}\dot{I}_{X_{2}}/\kappa$
for the symmetric reference state at fixed confinement
$\lambda=1$. The surface illustrates the dependence of the
scaled information flow on the nonequilibrium driving
($\delta$) and the dimensionless inertial
parameter ($\chi$). The solid red curve denotes the
optimal inertia ($\chi^{*}$), while the dashed curve on the
$(\delta/2,\chi)$ plane marks the upper boundary
($\chi_{\rm rng}$) beyond which inertia no longer enhances the
information flow relative to the overdamped limit. The red point
indicates the critical driving ($\delta_c/2$) at which the optimum
first enters the physical region ($\chi^{*}=0$), signaling the onset
of inertia-assisted information transfer. The blue point marks the
global maximum of the optimal inertia,
$(\delta_{\rm max}/2,\chi_{\rm max},
(\gamma_{0}\dot{I}_{X_{2}}/\kappa)_{\rm max})$. The vertical cliff near
$(\delta/2,\chi)=(1,0)$ highlights the singular corner where the limits
$\chi\rightarrow0$ and $\delta/2\rightarrow1$ do not commute,
demonstrating that the overdamped and finite-inertia descriptions
approach different asymptotic regimes.
}
\label{fig:UniversalGeometry}
\end{figure}

The optimal inertia follows from
${\partial\mathcal{F}}/{\partial\chi}=0$, with
the physical solution 
\begin{equation}
\chi^{*}
=
\frac{\delta\sqrt{4-\delta^2}}{4}
-
\frac{(1+\lambda)(4-\delta^2)}{4},
\label{eq:chiopt}
\end{equation}
which is positive only above the onset
$\chi^*=0$.
This gives the critical driving
\begin{equation}
\frac{\delta_c}{2}=
\frac{1+\lambda}{\sqrt{1+(1+\lambda)^2}}.
\end{equation}
Equivalently,
\begin{equation}
\lambda_c=
\frac{\delta}{\sqrt{4-\delta^2}}-1.
\end{equation}
Thus, inertia enhances information transfer only when $\delta>\delta_c$ or, equivalently, $\lambda<\lambda_c$.

The optimal inertia itself is non-monotonic. It increases from zero at $\delta_c/2$, reaches its maximum at
\begin{equation}
\frac{\delta_{\rm max}}{2}
=
\sqrt{\frac12\left(1+\frac{1+\lambda}{\sqrt{1+(1+\lambda)^2}}\right)},
\end{equation}
where
\begin{equation}
\chi_{\rm max}
=
\frac12\left(\sqrt{1+(1+\lambda)^2}-(1+\lambda)\right),
\end{equation}
and then decreases continuously to zero as $\delta/2\rightarrow1$.

Substituting eq.~(\ref{eq:chiopt}) into the scaling function yields the universal result
\begin{equation}
\left(\frac{\gamma_{0}}{\kappa}\dot I_{X_{2}}\right)_{\rm \chi^*}
=
\frac{1}{\sqrt{4-\delta^2}}, 
\end{equation}
which is independent of the confinement parameter $\lambda$. Thus, although the optimal inertia depends on both $\delta$ and $\lambda$, the maximum attainable information flow depends only on the nonequilibrium driving and increases monotonically with $\delta$. The value of the universal function at $\chi_{\rm max}$ is $\left(\gamma_{0} \dot{I}_{X_{2}}/\kappa \right)_{\rm max} = 1/\sqrt{4-\delta_{\rm max}^{2}}$.

Finally, while $\chi^*$ identifies the location of the maximum, the full range, over which inertia is beneficial, can be obtained from
\begin{equation}
\mathcal{F}(\delta,\chi_{\rm rng},\lambda)=\mathcal{F}(\delta,0,\lambda),
\end{equation}
which gives
\begin{equation}
\chi_{\rm rng}
=
\frac{1+(1+\lambda)^2}{4(1+\lambda)}
\left(\delta^2-\delta_c^2\right).
\end{equation}
Therefore, inertia enhances information transfer only within
$0<\chi<\chi_{\rm rng}$,
where $\chi_{\rm rng}=0$ at $\delta=\delta_c$ and increases monotonically with the nonequilibrium driving.

An intriguing feature of the universal scaling function is the singular behavior at the corner $(\delta/2,\chi)=(1,0)$, corresponding simultaneously to the extreme nonequilibrium limit ($T_2\rightarrow0$) and the overdamped limit ($M_{0}\rightarrow0$). In general, the limiting value of the information flow depends on the order in which these limits are taken. If the overdamped limit is taken first, the information flow approaches the finite value $\gamma_{0}\dot{I}_{X_{2}}/\kappa=1+\lambda$. In contrast, taking the extreme nonequilibrium limit while retaining a finite inertia gives $\gamma_{0}\dot{I}_{X_{2}}/\kappa\simeq1/\chi$, which diverges as $\chi\rightarrow0$. Along the optimal trajectory, where $2\chi^*\simeq\sqrt{4-\delta^2}$, the maximum information flow diverges more slowly as $\gamma_{0}\dot{I}_{X_{2}}/\kappa\simeq1/(\sqrt{4-\delta^2})\simeq1/2\chi^*$. Thus, the point $(\delta/2,\chi)=(1,0)$ is a singular point of the theory at which the overdamped and extreme nonequilibrium limits do not commute. Physically, this singularity reflects the competition between the inertial relaxation time and the increasingly asymmetric thermal driving as one reservoir approaches zero temperature. Consequently, infinitesimal inertia can qualitatively alter the asymptotic information transfer in the strongly driven regime, leading to dramatically different predictions depending on the experimental protocol by which the singular point is approached.

The universal scaling function derived above is a direct consequence of the complete symmetry of the reference system, where the two beads possess identical masses, friction coefficients, and trapping stiffnesses. In this highly symmetric limit, all microscopic parameters enter only through the single dimensionless inertial parameter $\chi$, leading to the universal geometry shown in Fig.~\ref{fig:UniversalGeometry}. Breaking any of these symmetries destroys this collapse, since the information flow can no longer be expressed solely in terms of $(\delta,\chi,\lambda)$. Instead, the response acquires additional independent parameter dependencies, leading to a richer multidimensional response landscape of information flow. While an exact analytical treatment rapidly becomes intractable, the behavior close to the symmetric reference state can be understood systematically through a perturbative expansion in the symmetry-breaking parameters. This perturbative framework, developed in the following subsection, quantifies how microscopic asymmetries deform the universal information landscape and identifies the leading mechanisms responsible for the differences we observe from the overdamped limit.

\subsection{Symmetry-breaking perturbations around the universal scaling regime}
\label{subsec:UD}
Having established the universal scaling function for the symmetric system, we now investigate how this response landscape is modified by weak microscopic asymmetries. To this end, we again parameterize the system as
\begin{equation}
m_{1,2}=M_{0}\left(1\pm\frac{\eta}{2}\right),
\end{equation}
along with eqs. (\ref{PerturbationParametersOD}) and (\ref{t12}). 

Since the above parametrization is introduced directly into the exact analytical expression for the information flow, $\delta$ can be considered as an independent control parameter but in our analysis, we have treated the perturbation and eigenvalues from a symmetric equilibrium reference state where $\delta=0$ for convenience.

As in the symmetric equal mass case, it is convenient to introduce the dimensionless inertial parameter ($\chi$) and obtain the dimensionless $\dot{\mathcal{I}}_{X_{2}}$ like from eqs.~(\ref{ScaledQuantities}) and (\ref{ScI_OD}) . Remarkably, after the above substitutions, the non-dimensionalized information flow again assumes the universal rational form
\begin{equation}
\dot{\mathcal{I}}_{X_{2}}
=
\frac{\mathcal{A}+\mathcal{B}\chi}
{\mathcal{C}+\mathcal{D}\chi+\mathcal{E}\chi^2},
\label{Univ_ScalForm_I}
\end{equation}
where the coefficients $\mathcal{A}$, $\mathcal{B}$, $\mathcal{C}$, $\mathcal{D}$, and $\mathcal{E}$ are functions of the dimensionless parameters $(\alpha,\delta,\epsilon,\eta,\lambda)$. This representation provides a convenient starting point for systematically investigating how weak microscopic asymmetries deform the universal information landscape obtained for the symmetric system.

\begin{figure}[t]
\centering
\includegraphics[width=\columnwidth]{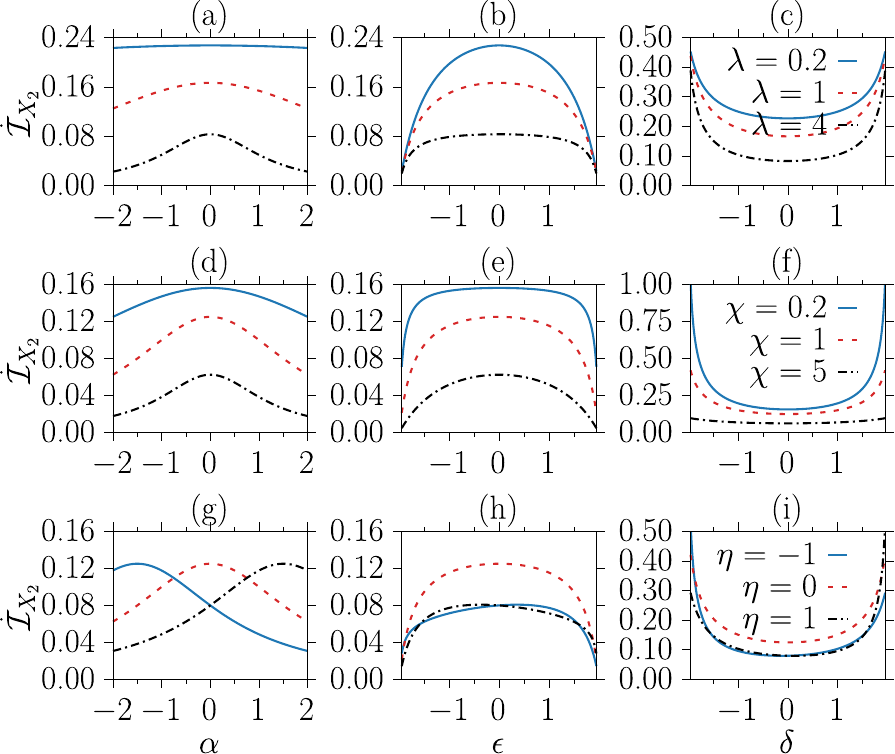}
\caption{Cross sections of the dimensionless information flow ($\dot{\mathcal{I}}_{X_{2}}$) illustrating the influence of weak system asymmetries. Panels (a)-(c) show the dependence on the stiffness asymmetry $\alpha$, friction asymmetry $\epsilon$, and normalized temperature difference $\delta$ for different values of the stiffness ratio $\lambda$ with $\chi=1$ and $\eta=0$. Panels (d)-(f) show the corresponding dependence for different inertia parameters $\chi$ at fixed $\lambda=2$ and $\eta=0$, while panels (g)-(i) illustrate the effect of mass asymmetry $\eta$ at fixed $\lambda=2$ and $\chi=1$. Unless varied, the parameters are fixed at $\delta=0$, $\alpha=0$, and $\epsilon=0$.}
\label{fig:Ibar_UD}
\end{figure}

Figure~\ref{fig:Ibar_UD} illustrates how inertial effects modify the perturbation landscape of the non-dimensionalized information flow $\dot{\mathcal{I}}_{X_{2}}$ obtained in the overdamped limit (Fig.~\ref{fig:PerturbationLandscape}). In contrast to the flat landscape observed in the overdamped case, the introduction of finite inertia induces a pronounced curvature in the information-flow landscape. As a result, the symmetric equilibrium configuration $(\alpha,\delta,\epsilon,\eta)=(0,0,0,0)$ becomes a local maximum with respect to both the stiffness asymmetry $\alpha$ and the friction asymmetry $\epsilon$, as evident from the cuts shown in Figs.~\ref{fig:Ibar_UD}(a), (b), (d), and (e). The dependence on the temperature asymmetry $\delta$, however, retains the same qualitative behavior as in the overdamped regime, increasing monotonically with increasing thermal bias [Figs.~\ref{fig:Ibar_UD}(c), (f)]. This nonlinear response of informational flow has a minimum at the symmetric equilibrium state with equal masses as can be seen from eq.~(\ref{ScI_cut_delta_UD}).
For a cut along $\delta$ direction, 
\begin{align}
\lambda_{\mathrm{bath}}(\mathbf{q})
&=
\frac{\kappa}
{2\gamma_{0}\left(
1+\lambda
+\chi
\right)},
\\[6pt]
\frac{\kappa}{\gamma_{0}}\mathcal{I}_{X_2}
&=
\lambda_{\mathrm{bath}}(\mathbf{q}) \times\nonumber\\
&
\left(
1+
\frac{
\left(\lambda^2+2\lambda+\lambda\chi+\chi\right)
}{
4(1+\lambda+\chi)^2
}\delta^2
+\mathcal{O}(\delta^3)
\right),
\label{ScI_cut_delta_UD}
\end{align}
\begin{align}
&\hspace{4cm}\text{where}\;
\mathbf{q}=(0,\delta,0,0).
\nonumber
\end{align}
The expression for $\lambda_{\mathrm{bath}}(\mathbf{q})$ could be derived from the expression for thermal bath conductivity in \cite{Thomas2026Kinetic} for underdamped dynamics.
A distinctive consequence of mass asymmetry is that the maximum of $\dot{\mathcal{I}}_{X_{2}}$ is displaced away from the symmetric reference state much pronounced with stiffness asymmetry ($\alpha$). This shift is observed clearly for the one-parameter cuts when varying the stiffness asymmetry $\alpha$ [Fig.~\ref{fig:Ibar_UD}(g)]  toward the direction in which the less stiffer spring is associated with the lighter particle. Furthermore, increasing the mass asymmetry reduces the value of $\dot{\mathcal{I}}_{X_{2}}$ at the symmetric reference state itself, as seen from the progressively lower values at $\alpha=\epsilon=\delta=0$ in figures in the third row of Fig.~\ref{fig:Ibar_UD}.

\subsection{Shift of the optimal stiffness asymmetry induced by mass asymmetry}

The perturbation analysis further reveals how inertia modifies the location of the optimal information flow. For a symmetric thermal and frictional environment ($\delta=\epsilon=0$), the condition for an extremum of the information flow with respect to the stiffness asymmetry,
\begin{equation}
\left.\frac{\partial \dot{\mathcal{I}}_{X_{2}}}{\partial \alpha}\right|_{\delta=\epsilon=0}=0,
\end{equation}
yields the simple analytical relation
\begin{equation}
\alpha_{\rm max}=\frac{1+\lambda}{\lambda}\eta.
\label{eq:alphamax}
\end{equation}

Equation~(\ref{eq:alphamax}) predicts that the location of the maximum of non-dimensionalized information flow varies linearly with the mass asymmetry. In particular, the symmetric reference state $(\alpha,\eta)=(0,0)$ ceases to be the point of maximum information transfer once the masses become unequal. Instead, the optimal stiffness asymmetry shifts in proportion to the mass asymmetry, with the slope determined solely by the dimensionless stiffness ratio $\lambda$.

This analytical prediction is consistent with the behavior observed in Fig.~\ref{fig:Ibar_UD}(g), where the maxima of the $\mathcal{I}_{X_{2}}(\alpha)$ curves move systematically away from the origin as $\eta$ is varied. A positive (negative) mass asymmetry shifts the maximum toward positive (negative) values of $\alpha$, indicating that the stiffness asymmetry required to maximize the information flow is directly coupled to the degree of mass asymmetry. Thus, unlike the overdamped limit, where the perturbation landscape remains flat with respect to $\alpha$, inertia introduces a preferred direction in parameter space and establishes a linear correspondence between the optimal stiffness and mass asymmetries.

\subsection{Constant-diffusion cut and emergence of bimodality}
\label{subsec:ConstD0}

To isolate the role of inertia and asymmetry in shaping the response landscape, we consider a special perturbation path defined by the constraint $\delta=\epsilon$ with $\alpha=0$. Along this line, the effective diffusion coefficients remain equal ($D_{1}=D_{2}\equiv D_0=T_{0}/\gamma_{0}$), thereby removing trivial asymmetries arising from unequal noise strengths. This constant-diffusion ($D_0$) cut provides a natural setting to probe the intrinsic effects of inertia, coupling, and mass asymmetry on the information flow.

\begin{figure}[t]
\centering
\includegraphics[width=0.9\linewidth]{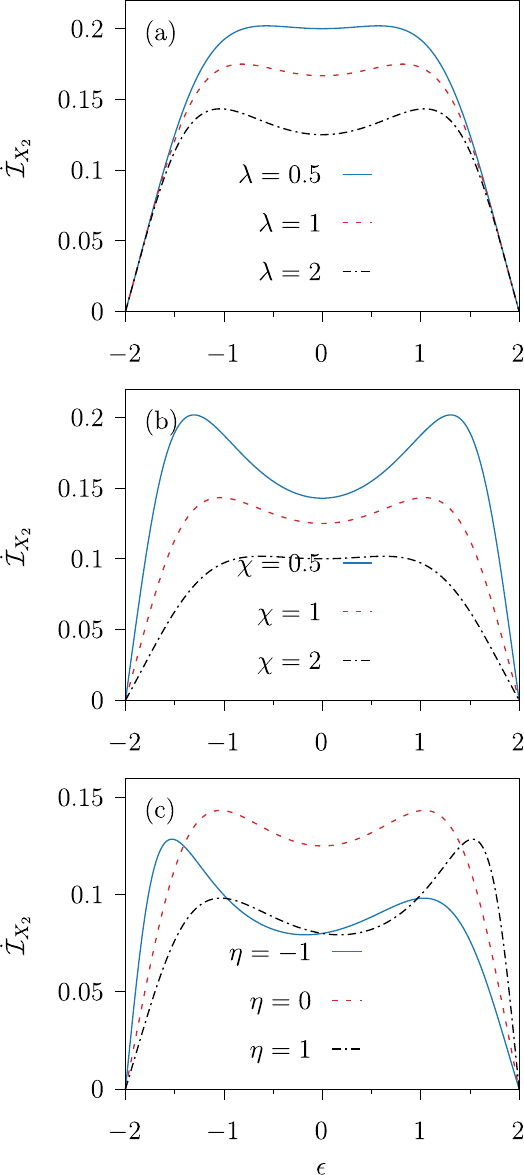}
\caption{
Dimensionless information flow ($\dot{\mathcal{I}}_{X_{2}}$) as a function of the frictional asymmetry parameter $\epsilon$ along the constant-diffusion cut $\delta=\epsilon$ with $\alpha=0$. 
(a) Variation with coupling asymmetry $\lambda$ for fixed $\chi=1$ and $\eta=0$. 
(b) Variation with inertial parameter $\chi$ for fixed $\lambda=2$ and $\eta=0$. 
(c) Variation with mass asymmetry $\eta$ for fixed $\lambda=2$ and $\chi=1$.}
\label{fig:Ibar_eps_cut}
\end{figure}

Figure~\ref{fig:Ibar_eps_cut} shows the behavior of the dimensionless information flow $\dot{\mathcal{I}}_{X_{2}}$ as a function of $\epsilon$ along this cut. In the overdamped limit, the corresponding landscape is a normal minima at origin and increasing symmetrically away from it. In contrast, the inclusion of inertia leads to the emergence of nontrivial structure, including curvature and, notably, bimodality.

As shown in Fig.~\ref{fig:Ibar_eps_cut}(a), increasing the coupling parameter $\lambda$ induces the development of a bimodal structure in $\dot{\mathcal{I}}_{X_{2}}(\epsilon)$, with two symmetric peaks emerging away from $\epsilon=0$. Similarly, Fig.~\ref{fig:Ibar_eps_cut}(b) demonstrates that decreasing the inertial parameter $\chi$ enhances this effect, driving the system from a unimodal to a bimodal regime. Physically, both stronger coupling and reduced effective inertia amplify correlations between the degrees of freedom, thereby favoring the formation of distinct extrema in the response landscape.

Mass asymmetry introduces an additional directional bias. As seen in Fig.~\ref{fig:Ibar_eps_cut}(c), increasing $\eta$ not only modifies the curvature but also enhances the peak in the direction where the lighter particle experiences a smaller frictional drag coefficient. This reflects the fact that the lighter, less damped particle responds more strongly to fluctuations, thereby dominating the information transfer in that direction.

The emergence of bimodality allows us to define a phase diagram in the $(\chi,\lambda)$ plane. On one side of the phase boundary, $\dot{\mathcal{I}}_{X_{2}}(\epsilon)$ exhibits a single maximum at $\epsilon=0$, corresponding to a unimodal response landscape. On the other side, two symmetric maxima appear, signaling a bimodal structure. The phase boundary is determined by the condition that the curvature at the origin changes sign, i.e.,
\begin{equation}
\left.
\frac{d^2 \dot{\mathcal {I}}_{X_{2}}(0,\epsilon,\epsilon,0)}{d\epsilon^2}
\right|_{\epsilon=0}
= 0,
\label{eq:curvature_condition}
\end{equation}

Using the analytical scaling form (eq.~(\ref{Univ_ScalForm_I}))
and expanding around $\epsilon=0$ along $\delta=\epsilon$, the curvature can be evaluated explicitly. The transition from unimodal to bimodal behavior occurs when the quadratic coefficient in the expansion vanishes. This condition yields the critical line
\begin{equation}
\chi^{pb}(\lambda)
=
\sqrt{\frac{1+4\lambda+2\lambda^2}{2}}
\label{eq:phase_boundary}
\end{equation}

which defines the phase boundary separating the two regimes.

\begin{figure}
\centering
\includegraphics[width=0.9\columnwidth]{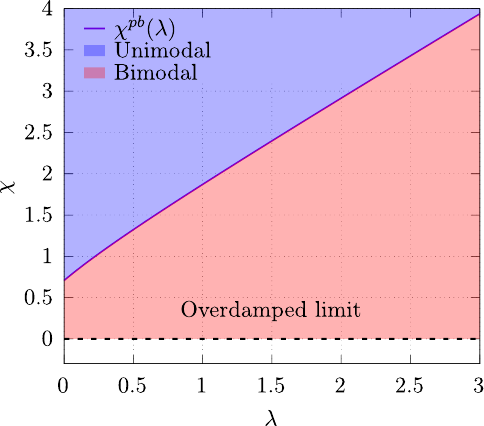}
\caption{
Phase diagram in the $(\chi,\lambda)$ plane along the constant diffusion
manifold $\delta=\epsilon$. The solid curve
$\chi^{pb}(\lambda)$ marks the boundary
obtained from the curvature condition to change sign. The response landscape of information flow exhibits a single maximum (unimodal) in the violet region, while in the red region it develops two symmetric maxima (bimodal). The overdamped limit
lies within the bimodal region.}
\label{fig:phase_diagram}
\end{figure}

To summarize the emergence of bimodality, we construct a phase diagram
in the $(\chi,\lambda)$ plane along the constant diffusion path
$\delta=\epsilon$ in Fig. (\ref{fig:phase_diagram}). This boundary separates a region where the response landscape is
characterized by a single peak from a region where it develops two
distinct maxima, indicating the configuration with non-zero symmetry breaking parameters having higher response to informational flow than symmetrical reference state. Interestingly, the
overdamped limit lies entirely within the bimodal region, suggesting
that the bimodality is already present at the level of overdamped
dynamics, while inertia influences the manner in which this structure
is realized in the underdamped system.

Thus, the constant-diffusion cut reveals that inertia does not merely deform the overdamped landscape but qualitatively reorganizes it, leading to symmetry breaking and the emergence of bimodality. The resulting phase diagram provides a compact characterization of how coupling strength and inertia compete to shape nonequilibrium information transfer.

\subsection{Principal-curvature analysis of the underdamped response landscape}
\label{sec:HessianAnalysis}

To characterize how inertia modifies the local geometry of the information
landscape, we examine the Hessian of the dimensionless information flow
$\dot{\mathcal I}_{X_{2}}$ at the symmetric reference state,
\begin{equation}
\mathbf{q}=(\alpha,\delta,\epsilon,\eta)=(0,0,0,0).
\end{equation}
The Hessian is defined as
\begin{equation}
H_{ij}
=
\left.
\frac{\partial^2 \dot{\mathcal{I}}_{X_{2}}}
{\partial q_i\partial q_j}
\right|_{\mathbf{q}=\mathbf{0}},
\qquad
\mathbf{q}=(\alpha,\delta,\epsilon,\eta),
\end{equation}
so that, locally,
\begin{equation}
\dot{\mathcal{I}}_{X_{2}}(\mathbf{q})
\simeq
\dot{\mathcal{I}}_{X_{2}}(\mathbf{0})
+
\frac{1}{2}\mathbf{q}^{T}H\mathbf{q}.
\label{eq:HessianExpansion}
\end{equation}
The eigenvalues of $H$ therefore determine the principal curvatures
of the response landscape, while the corresponding eigenvectors
identify the collective perturbation directions along which these
curvatures act. 

\begin{figure}[t]
    \centering
    \includegraphics[width=\linewidth]{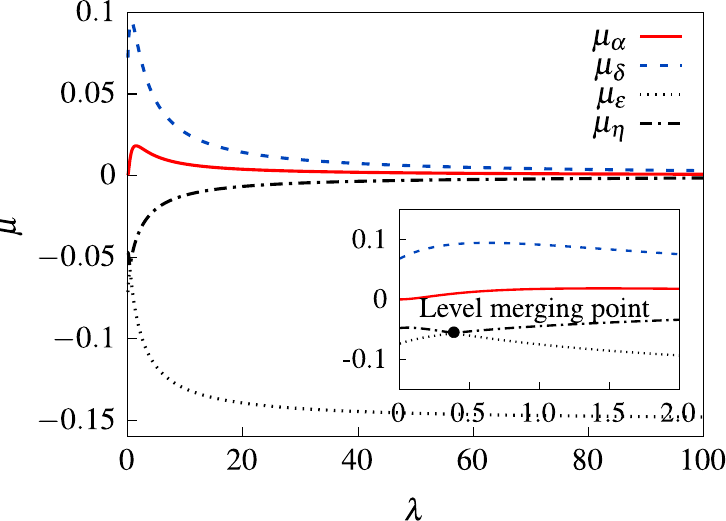}
    \caption{Eigenvalues of the Hessian of the dimensionless information-flow
    landscape at the symmetric reference state as a function of $\lambda$ for
    $\chi_{m}\simeq0.3$. The four branches correspond to the principal modes dominated
    by the $\alpha$, $\delta$, $\epsilon$, and $\eta$ perturbations,
    respectively. The inset magnifies the small-$\lambda$ regime, where the
    $\epsilon$- and $\eta$-dominated eigenvalue branches approach and touch at
    $(\lambda_{m},\mu_{m})\simeq(0.39,-0.055)$, indicated by the black marker.}
    \label{fig:HessianEigenvalues}
\end{figure}

The resulting evolution of the principal curvatures with $\lambda$ is
shown in Fig.~\ref{fig:HessianEigenvalues}. In the overdamped limit, the
curvature associated predominantly with the frictional asymmetry
$\epsilon$ approaches zero at large $\lambda$, corresponding to the
asymptotically flat perturbation landscape observed in the one-dimensional
cuts. Finite inertia qualitatively modifies this behavior. The
$\alpha$- and $\delta$-dominated curvatures remain positive and decrease
towards zero with increasing $\lambda$, whereas the curvature associated
predominantly with the mass asymmetry, represented by $\eta$ remains negative and
approaches a finite negative value. The friction-asymmetry mode, is negatively curved at finite inertia but becomes
progressively softer with increasing $\lambda$, with its curvature
approaching zero in the strong-confinement limit. Thus, at finite
inertia, the symmetric reference state is locally a saddle point of
the response landscape of information flow, with two positive- and two negative-curvature
principal directions. In the large-$\lambda$ limit, the landscape
develops asymptotically soft directions while retaining a finite
negative curvature along the mass-asymmetry mode.


The two principal modes associated predominantly with $\epsilon$ and
$\eta$ also exhibit a level merging at intermediate values of $\chi$
and $\lambda$, as shown in inset of Fig.~\ref{fig:HessianEigenvalues}. The two
eigenvalue branches form an $X$-shaped spectrum and become degenerate
at the merging. At the degeneracy, the corresponding two-dimensional
eigenspace is not uniquely resolved into individual principal
directions, and arbitrary orthogonal combinations within this subspace
are equally valid eigenvectors. Away from the degeneracy, the two
eigenvalues separate and the corresponding eigenvectors evolve
continuously, with their relative contributions from frictional and
mass asymmetries changing with the inertial parameter. The level
merging therefore represents a reorganization of the principal
perturbation modes rather than a discontinuity in the information flow
itself. The numerically obtained level merging point is ($\lambda_{m}=0.3993\pm0.0001,\chi_{m}=2885\pm0.0001$).

It is important to distinguish this level merging from the
unimodal--bimodal transition identified from the constant-$D_0$ cut.
The latter is associated with a change in the curvature of the
response landscape along a physically selected perturbation path,
whereas the former concerns a degeneracy of the local principal
curvatures at the symmetric reference state. Together, these results
show that the underdamped response landscape of information flow contains geometric
features absent in the overdamped limit: inertia curves the previously
flat landscape, produces a saddle structure with distinct positive and
negative curvature directions, introduces a unstable mass-asymmetry mode,
and induces a degeneracy between the frictional- and mass-asymmetry
principal modes at intermediate inertia.

\subsection{Symmetry relations under mass exchange}
\label{app:SRME}
We now derive a set of symmetry relations for the information flow and
entropy production under exchange of particle masses
$(m_1,m_2)\leftrightarrow(m_2,m_1)$ for the symmetric system
($k_1=k_2$, $\gamma_1=\gamma_2$).

From Ref. \cite{Thomas2026Kinetic}, the total entropy production is invariant under
mass exchange,
\begin{equation}
\dot{\Sigma}(m_1,m_2)=\dot{\Sigma}(m_2,m_1),
\end{equation}
and the steady-state heat currents satisfy
\begin{align}
\dot{Q}_{X_{1}}(m_1,m_2)=\dot{Q}_{X_{1}}(m_2,m_1)\\
\dot{Q}_{X_{2}}(m_1,m_2)=\dot{Q}_{X_{2}}(m_2,m_1).
\end{align}
We further use the decomposition
\begin{equation}
\dot{\Sigma}_{X_{1}}=-\frac{\dot{Q}_{X_{1}}}{T_1}-\dot{I}_{X_{1}},
\qquad
\dot{\Sigma}_{X_{2}}=-\frac{\dot{Q}_{X_{2}}}{T_2}-\dot{I}_{X_{2}},
\end{equation}
together with the steady-state condition
\begin{equation}
\dot{I}_{X_{1}}=-\dot{I}_{X_{2}}.
\end{equation}

By definition of the bipartite information flow, the combination
\begin{equation}
\dot{I}_{X_{1}}(m_1,m_2)+\dot{I}_{X_{1}}(m_2,m_1)
\end{equation}
is invariant under mass exchange. Using
$\dot{I}_{X_{1}}(m_2,m_1)=-\dot{I}_{X_{2}}(m_2,m_1)$, this immediately implies that
\begin{equation}
\dot{I}_{X_{1}}(m_1,m_2)-\dot{I}_{X_{2}}(m_2,m_1)
\end{equation}
is symmetric under mass exchange, while the sum
\begin{equation}
\dot{I}_{X_{1}}(m_1,m_2)+\dot{I}_{X_{2}}(m_2,m_1)
\end{equation}
is anti-symmetric.

Substituting the above relations into the entropy production
decomposition shows that
\begin{equation}
\dot{\Sigma}_{X_{1}}(m_1,m_2)-\dot{\Sigma}_{X_{2}}(m_2,m_1)
\end{equation}
is symmetric under mass exchange, consistent with the invariance of the
total entropy production. In contrast, the individual components
$\dot{\Sigma}_{X_{1}}$ and $\dot{\Sigma}_{X_{2}}$ are not invariant, reflecting the
redistribution of dissipation between the two reservoirs.

These symmetry relations demonstrate that while the total entropy
production and heat currents are insensitive to mass permutation, the
information flow and the partitioning of entropy production retain a
nontrivial dependence on the mass distribution. This highlights the
distinct role of inertia in controlling the directionality of
information transfer without altering the overall irreversibility of
the system.

\section{Discussion}
\label{sec:Discussion}
A useful reference state for interpreting the information-flow landscape is $T_1=T_2=T_0$, or $\delta=0$. Although
the information flow vanishes at this point, but the thermal conductivity for this state depends on the remaining
asymmetry parameters which characterizes the zero order coefficient of response to perturbations of information flow away from
the symmetric state. The maxima and minima in the perturbation cuts therefore
identify parameter regimes of enhanced and suppressed response of information transfer. In particular, for underdamped dynamics, the displacement of the maxima with mass
asymmetry  shows that inertia can select a
preferred asymmetric configuration for information transfer.

The equilibrium reference point also provides a natural starting point for
weakly nonequilibrium states. For small $\delta$, corresponding to the
linear-response regime, the qualitative structure of the perturbation
landscape remains largely unchanged. Extending the Hessian analysis to finite
$\delta$ could reveal how the principal curvatures and eigenvectors evolve
toward the onset $\delta_c$ of information-flow enhancement found in the
underdamped limit.

A complementary direction is to analyze the underdamped entropy-production
rates $\dot{\Sigma}_{X_{i}}$. Their covariance-matrix
representation provides a framework for quantifying how information flow,
heat transfer, and dissipation are partitioned between the two reservoirs
and how this partition is modified by inertia and microscopic asymmetries.

In the present autonomous setup, information is generated but not actively harnessed for power generation. A feedback or state-dependent jump protocol that modifies the coupling potential could provide a route to convert this information into mechanical work, in the spirit of bipartite information engines and Maxwell-demon mechanisms \cite{HorowitzEsposito2014,SagawaUeda2010,EspositoSchaller2012}. In the
present model, the work associated with such a jump would depend on the
instantaneous coupling force and configuration, providing a natural
extension toward information-to-work conversion. This framework may also
be relevant to molecular machines such as kinesin, where information and
energy flows have been analyzed within a bipartite stochastic description
\cite{Leighton2024,DuBuisson2025}. Such an extension could test whether the information-flow enhancement at large thermal asymmetry along specific paths in the ($\delta/2$,$\chi$) plane can be exploited for work extraction. It could also examine whether the mass-asymmetry-induced shift of the response of information-flow maxima provides a route to optimize work extraction in the underdamped system.

An intriguing feature of the underdamped dynamics is the divergence of the information flow observed at $\delta/2=1$ and $\chi=0$ when the masses are reduced toward zero along the corresponding inertial path. Within the
present classical Langevin description, decreasing the mass increases the
characteristic velocity fluctuations, and the resulting divergence signals a singular limit of the model rather than a physically realizable infinite information-transfer rate. In this regime, additional microscopic physics may become relevant. In particular, sufficiently small masses can require a quantum description of Brownian motion, while sufficiently large thermal velocity fluctuations can invalidate the nonrelativistic approximation and require relativistic stochastic dynamics \cite{CaldeiraLeggett1983,DunkelHanggi2009}. Relativistic Brownian-motion theories explicitly incorporate the bounded nature of particle velocities and the corresponding modifications of Langevin and Fokker--Planck dynamics \cite{DunkelHanggi2009}. Thus, the divergence found here may be viewed as an indication of the boundary of validity of the classical nonrelativistic model and as a possible entry point for investigating quantum or relativistic corrections to information-flow generation.

Beyond the magnitude of information flow, we investigate its response to
microscopic symmetry breaking using a dimensionless information-flow
landscape. The overdamped landscape contains flat directions, while inertia
curves this landscape, shifts the maxima with mass asymmetry, and produces
bimodality and level touching of the principal curvatures in the underdamped
regime.


The structure of the information-flow landscape also suggests a connection
with Landau-type descriptions of nonequilibrium phenomena. Although
$\dot{\bar I}_{X_{2}}$ is not a free energy and its extrema should not be
identified with equilibrium phases, its landscape provides an analogous
geometric description in terms of preferred responses, curvature, and
bimodality. The emergence of two maxima along the constant-$D_{0}$
cut, together with their systematic displacement under mass asymmetry, further enhances the connection of the geometric analogy of informational flow with Landau-type effective landscapes. This motivates a more systematic nonequilibrium framework in which an information-flow functional, together with a suitable order parameter and large-deviation description, could characterize competing information-generating states.
\cite{MeibohmEsposito2022,MeibohmEsposito2023}. The present two-bead model may serve as a minimal framework for exploring this idea, with extensions to
larger heterogeneous networks and dynamical environments offering a
possible route toward a broader theory of information landscapes in
nonequilibrium systems.

\appendix

\section{Information flow from the covariance matrix}
\label{app:InformationFlow}

For linear Langevin systems, the steady-state probability distribution is
Gaussian and is therefore completely characterized by the covariance matrix \cite{van2004microscopic, seifert2012stochastic, barato2015thermodynamic}.
This property allows the information flow to be evaluated analytically without
explicitly solving the Fokker--Planck equation.

The information flow into subsystem $X_2$ is defined as~\cite{HorowitzEsposito2014,Hartich2014} as 
\begin{equation}
\dot I_{X_{2}}
=
\int d{\bf z}\,
J_{X_{2}}({\bf z})
\cdot
\nabla_{X_{2}}
\ln P(X_{1}|X_{2}),
\label{eq:IdotDef}
\end{equation}
where
${\bf z}$ denotes the complete phase-space vector,
$J_{X_{2}}$ is the probability current associated with subsystem $X_{2}$, and
$P(X_{1}|X_{2})$ is the conditional probability density.

For the two-bead system, we partition the phase-space vector as
\begin{equation}
{\bf z}
=
\begin{pmatrix}
{\bf z}_1\\
{\bf z}_2
\end{pmatrix},
\qquad
{\bf z}_1=(x_1,v_1)^T,
\qquad
{\bf z}_2=(x_2,v_2)^T .
\end{equation}

The steady-state distribution is a four-dimensional Gaussian,
\begin{equation}
P({\bf z})
=
\frac{1}
{(2\pi)^{2}\sqrt{\det\sigma}}
\exp
\left(
-\frac{
{\bf z}^T
\sigma^{-1}
{\bf z}}{2}
\right),
\end{equation}
where the Gaussian form follows from the Ornstein--Uhlenbeck nature of the Langevin process.~\cite{Risken1989,Gardiner2009}. The covariance matrix is written in block form as
\begin{equation}
\sigma
=
\begin{pmatrix}
\sigma_{11} & \sigma_{12}\\
\sigma_{21} & \sigma_{22}
\end{pmatrix}.
\end{equation}

Since every marginal of a Gaussian distribution is itself a Gaussian function, the
conditional probability also has a Gaussian form,
\begin{align}
P({\bf z}_2&|{\bf z}_1)
=\frac{\exp
\left(
-{
({\bf z}_2-\boldsymbol{\mu})^T
\sigma_c^{-1}
({\bf z}_2-\boldsymbol{\mu})}/{2}
\right)}{2\pi\sqrt{\det\sigma_c}},
\end{align}
where 
$\boldsymbol{\mu}
=
\sigma_{21}
\sigma_{11}^{-1}
{\bf z}_1$, and
$\sigma_c
=
\sigma_{22}
-
\sigma_{21}
\sigma_{11}^{-1}
\sigma_{12}$.
Taking the logarithmic derivative of eq. (A5) yields 
\begin{equation}
\nabla_{X_{2}}
\ln P({\bf z}_2|{\bf z}_1)
=
-
\sigma_c^{-1}
\left(
{\bf z}_2-\boldsymbol{\mu}
\right).
\label{eq:ConditionalGradient}
\end{equation}

For the underdamped dynamics, the probability current is
\begin{equation}
{\bf J}
=
\left(
A
+
D\sigma^{-1}
\right)
{\bf z}\,
P({\bf z})={\bf j}P({\bf z}),
\end{equation}
obtained from the stationary Fokker--Planck equation  \cite{Risken1989,Gardiner2009}, where $A$ is the drift matrix and $D$ is the diffusion matrix obtained from the Langevin equation ($\mathbf{\dot z} = A\bf z + \boldsymbol{\xi}$) and $\langle \boldsymbol{\xi}(t)\boldsymbol{\xi}^{T}(t')\rangle = 2D\delta(t-t')$ as written in eqs.~(\ref{pos1_Langevin}-\ref{vel2_Langevin}). The ${\bf j}$ is the probability velocity.
The current associated with subsystem $X_2$ is obtained by selecting the
rows corresponding to $(x_2,v_2)$ which can be mapped to the corresponding component of probability velocity,
\begin{equation}
{\bf J}_2
=
\left(
A
+
D\sigma^{-1}
\right)_{(3,4)}
{\bf z}\,
P({\bf z}) = {\bf j}_{2}P({\bf z}).
\end{equation}

Substituting eqs.~(\ref{eq:ConditionalGradient}) and
(\ref{eq:IdotDef}) gives
\begin{equation}
\dot I_{X_{2}}
=
-
\Big\langle
{\bf j}_2
\cdot
\sigma_c^{-1}
({\bf z}_2-\boldsymbol{\mu})
\Big\rangle .
\label{eq:IdotAverage}
\end{equation}
Since both the probability current and the conditional score are linear
functions of the phase-space variables, the integrand is quadratic in
${\bf z}$. The Gaussian average therefore reduces entirely to second
moments,
\begin{equation}
\langle
z_i z_j
\rangle
=
\sigma_{ij},
\end{equation}
so that the information flow can be expressed solely in terms of the
elements of the covariance matrix.

The resulting expression may be written compactly as
\begin{equation}
\boxed{
\dot I_{X_{2}}
=
-
\left\langle
{\bf j}_2
\cdot
\sigma_c^{-1}
({\bf z}_2-\sigma_{21}\sigma_{11}^{-1}{\bf z}_1)
\right\rangle ,
}
\label{eq:IdotCovariance}
\end{equation}
where every average is evaluated using the covariance matrix
$\sigma$. Substituting the steady-state covariance matrix, obtained from the Lyapunov equation
\begin{equation}
A\sigma+\sigma A^T=2D,
\label{eq:Lyapunov_eqn}
\end{equation}
for the stationary multivariate Ornstein--Uhlenbeck process \cite{Gardiner2009,Risken1989}, and using its analytical solution from Ref.~\cite{Thomas2026Kinetic}, gives the exact closed-form expression for the underdamped information flow presented in Eq.~\eqref{IdotY_UD}.

The overdamped expression follows identically by replacing
${\bf z}_i=(x_i)$, reducing the covariance matrix to its positional
subspace and evaluating the corresponding conditional 
Gaussian distribution.

\section{Overdamped dynamics}
\label{app:Covariance}
In this regime, the Langevin equation in eqs.~(\ref{pos1_OD_langevin}-\ref{pos2_OD_langevin}) may be written as
\begin{equation}
\dot{\mathbf{x}}
=
A\mathbf{x}
+\boldsymbol{\xi},
\end{equation}
where
\begin{equation}
\mathbf{x}
=
\begin{pmatrix}
x\\
y
\end{pmatrix},
\qquad
\langle
\boldsymbol{\xi}(t)
\boldsymbol{\xi}^{T}(t')
\rangle
=
2D\delta(t-t').
\end{equation}

For the two-bead system,
\begin{equation}
A=
\begin{pmatrix}
-\dfrac{k_1+\kappa}{\gamma_1}
&
\dfrac{\kappa}{\gamma_1}
\\[3mm]
\dfrac{k_2}{\gamma_2}
&
-\dfrac{k_2+\kappa}{\gamma_2}
\end{pmatrix},
\;
D=
\begin{pmatrix}
{T_{1}}/{\gamma_{1}}&0\\
0& {T_{2}}/{\gamma_{2}}
\end{pmatrix}.
\end{equation}
Since the dynamics is linear, the steady-state distribution is again Gaussian.
\begin{equation}
P(\mathbf{x})
=
\frac{1}
{2\pi\sqrt{\det\sigma}}
\exp
\left(
-\frac{
\mathbf{x}^{T}
\sigma^{-1}
\mathbf{x}}{2}
\right),
\end{equation}
where the covariance matrix
\begin{equation}
\sigma=
\begin{pmatrix}
S_{11}&S_{12}\\
S_{21}&S_{22}
\end{pmatrix}
\end{equation}
is obtained from the continuous Lyapunov equation (eq.~(\ref{eq:Lyapunov_eqn})).
The analytical solution of eq.~(\ref{eq:Lyapunov_eqn}) for the covariance matrix of the overdamped two-bead system has been taken from Ref.~\cite{Thomas2026Kinetic}.

The inverse covariance matrix is
\begin{equation}
\sigma^{-1}
=
\frac{1}{\Omega}
\begin{pmatrix}
S_{22}&-S_{12}\\
-S_{21}&S_{11}
\end{pmatrix},
\end{equation}
where
$\Omega =S_{11}S_{22}-S_{12}^{2}$.
The probability current is
\begin{equation}
\mathbf{J}
=
\left(
A+D\sigma^{-1}
\right)
\mathbf{x}
P(\mathbf{x}).
\end{equation}

Defining the drift vectors
\begin{align}
\mathbf{b}_{X_{1}}
&=
\left(
-\frac{k_1+\kappa}{\gamma_1},
\frac{\kappa}{\gamma_1}
\right),
\\
\mathbf{b}_{X_{2}}
&=
\left(
\frac{\kappa}{\gamma_2},
-\frac{k_2+\kappa}{\gamma_2}
\right),
\end{align}
the corresponding current coefficients are
\begin{align}
\mathbf{A}_{X_{1}}
&=
\mathbf{b}_{X_{1}}
+
D_{1}
\left(
\sigma^{-1}
\right)_1,
\\
\mathbf{A}_{X_{2}}
&=
\mathbf{b}_{X_{2}}
+
D_{2}
\left(
\sigma^{-1}
\right)_2,
\end{align}
where $(\sigma^{-1})_i$ denotes the $i$th row of the inverse covariance matrix.

The local entropy production rates associated with each reservoir follow directly from the Gaussian average of the irreversible current,
\begin{align}
\dot{\Sigma}_{X_{1}}
&=
\frac{1}{D_1}
\mathbf{A}_{X_{1}}
\sigma
\mathbf{A}_{X_{1}}^{T},\label{Sigma_X1}
\\
\dot{\Sigma}_{X_{2}}
&=
\frac{1}{D_2}
\mathbf{A}_{X_{2}}
\sigma
\mathbf{A}_{X_{2}}^{T},\label{Sigma_X2}
\end{align}
and the total entropy production rate is
\begin{equation}
\dot{\Sigma}
=
\dot{\Sigma}_{X_{1}}+\dot{\Sigma}_{X_{2}}.
\end{equation}
The covariance matrix representation of the local entropy production rates follows directly from the Gaussian current formalism and was previously derived for the overdamped model in Ref.~\cite{Thomas2026Kinetic}.
The information flow is obtained from the conditional probability
\begin{equation}
P(x|y)
=
\frac{P(x,y)}{P(y)},
\end{equation}
which, for a Gaussian distribution, becomes
\begin{equation}
P(x|y)
=
\frac{1}
{2\pi\sqrt{(S_{11}-S_{12}^{2}/S_{22})}}
\exp
\left(
-\frac{
\left(
x-\dfrac{S_{12}}{S_{22}}y
\right)^2
}
{2(S_{11}-S_{12}^{2}/S_{22})}
\right).
\end{equation}
The conditional Gaussian distribution and its moments follow from standard multivariate Gaussian identities~\cite{Bishop2006,Petersen2012}.

Taking the logarithmic derivative yields
\begin{equation}
\frac{\partial}{\partial y}
\ln P(x|y)
=
\frac{S_{12}}
{S_{11}S_{22}-S_{12}^{2}}
\left(
x-
\frac{S_{11}}{S_{12}}y
\right).
\end{equation}

Using the probability current and velocity
\begin{equation}
J_{X_{2}}
=
(\mathbf{A}_{X_{2}}\cdot\mathbf{x})P = j_{X_{2}}P,
\end{equation}
Following Ref. \cite{HorowitzEsposito2014}, the information flow is
\begin{equation}
\dot I_{X_{2}}
=
\left\langle
j_{X_{2}}
\frac{\partial}{\partial y}
\ln P(x|y)
\right\rangle .
\end{equation}

Since the integrand is quadratic in the Gaussian variables, all averages reduce to second moments,
\begin{equation}
\langle x^{2}\rangle=S_{11},
\qquad
\langle y^{2}\rangle=S_{22},
\qquad
\langle xy\rangle=S_{12},
\end{equation}
leading to the compact covariance representation
\begin{equation}
\boxed{
\dot I_{X_{2}}
=
\frac{S_{12}}{S_{22}}
\left(
\frac{k_2}{\gamma_2}
-
D_{2}
\frac{S_{12}}
{S_{11}S_{22}-S_{12}^{2}}
\right).
}
\label{eq:IdotCovOD}
\end{equation}
Using the analytical covariance matrix of Ref.~\cite{Thomas2026Kinetic}, the information flow reduces to the compact expressions as shown in eq. (\ref{IdotY_OD}).

Similarly, the information flow into subsystem $X_{2}$ from $X_{1}$ is
\begin{equation}
\boxed{
\dot I_{X_{1}}
=
\frac{S_{12}}{S_{11}}
\left(
\frac{k_2}{\gamma_1}
-
D_{1}
\frac{S_{12}}
{S_{11}S_{22}-S_{12}^{2}}
\right).
}
\label{eq:IdotCovODX}
\end{equation}

Substituting the steady-state covariance matrix into Eqs.~(\ref{eq:IdotCovOD}) and (\ref{eq:IdotCovODX}) yields
\begin{equation}
\dot I_{X_{1}}+\dot I_{X_{2}}=0,
\end{equation}
in agreement with the general identity
\begin{equation}
\frac{dI(X_{1}:X_{2})}{dt}=0
\end{equation}
derived in Ref. \cite{HorowitzEsposito2014}. Since the mutual information, $I(X_{1}:X_{2})$, is stationary in the non-equilibrium steady state, implying that the information flow out of one subsystem is exactly balanced by that into the other.

Equations~(\ref{eq:Lyapunov_eqn}), (\ref{Sigma_X1}), (\ref{Sigma_X2}) and (\ref{eq:IdotCovODX}) provide a complete covariance matrix formulation of the entropy production and information flow for linear overdamped Langevin systems. The underdamped expressions presented in the main text are obtained by following the same procedure after enlarging the state vector to include the velocity degrees of freedom.

\bibliography{InformationRate_TwoBeadSystem.bib}

@article{Landauer1961,
  author = {R. Landauer},
  title = {Irreversibility and Heat Generation in the Computing Process},
  journal = {IBM Journal of Research and Development},
  volume = {5},
  pages = {183--191},
  year = {1961}
}

@article{Bennett1982,
  author = {C. H. Bennett},
  title = {The Thermodynamics of Computation—A Review},
  journal = {International Journal of Theoretical Physics},
  volume = {21},
  pages = {905--940},
  year = {1982}
}

@article{Seifert2012,
  author = {U. Seifert},
  title = {Stochastic Thermodynamics, Fluctuation Theorems and Molecular Machines},
  journal = {Reports on Progress in Physics},
  volume = {75},
  pages = {126001},
  year = {2012}
}

@article{HorowitzEsposito2014,
  author = {Jordan M. Horowitz and Massimiliano Esposito},
  title = {Thermodynamics with Continuous Information Flow},
  journal = {Physical Review X},
  volume = {4},
  pages = {031015},
  year = {2014}
}

@article{HartichBaratoSeifert2014,
  author = {David Hartich and Andre C. Barato and Udo Seifert},
  title = {Stochastic Thermodynamics of Bipartite Systems: Transfer Entropy Inequalities and a Maxwell's Demon Interpretation},
  journal = {Journal of Statistical Mechanics},
  pages = {P02016},
  year = {2014}
}

@article{Parrondo2015,
  author = {J. M. R. Parrondo and J. M. Horowitz and T. Sagawa},
  title = {Thermodynamics of Information},
  journal = {Nature Physics},
  volume = {11},
  pages = {131--139},
  year = {2015}
}

@article{DuBuisson2025,
  author = {Johan du Buisson and Jannik Ehrich and Matthew P. Leighton and Avijit Kundu and Tushar K. Saha and John Bechhoefer and David A. Sivak},
  title = {Hunting for Maxwell's Demon in the Wild},
  journal = {arXiv:2504.11329},
  year = {2025}
}

@article{BaratoSeifert2014,
  author = {Andre C. Barato and Udo Seifert},
  title = {Unifying Three Perspectives on Information Processing in Stochastic Thermodynamics},
  journal = {Physical Review Letters},
  volume = {112},
  pages = {090601},
  year = {2014}
}

@article{Barato2014,
  author = {Barato, Andr\'e C. and Hartich, David and Seifert, Udo},
  title = {Information-Theoretic versus Thermodynamic Entropy Production in Autonomous Sensory Networks},
  journal = {Physical Review E},
  volume = {89},
  pages = {042104},
  year = {2014}
}

@article{Mandal2012,
  author = {Mandal, Dibyendu and Jarzynski, Christopher},
  title = {Work and Information Processing in a Solvable Model of Maxwell's Demon},
  journal = {Proceedings of the National Academy of Sciences},
  volume = {109},
  pages = {11641--11645},
  year = {2012}
}

@article{saito2007fluctuation,
  title={Fluctuation theorem in quantum heat conduction},
  author={Saito, Keiji and Dhar, Abhishek},
  journal={Physical Review Letters},
  volume={99},
  number={18},
  pages={180601},
  year={2007},
  publisher={APS}
}

@article{kundu2011large,
  title={Large deviations of heat flow in harmonic chains},
  author={Kundu, Anupam and Sabhapandit, Sanjib and Dhar, Abhishek},
  journal={Journal of Statistical Mechanics: Theory and Experiment},
  volume={2011},
  number={03},
  pages={P03007},
  year={2011},
  publisher={IOP Publishing}
}

@article{fogedby2012heat,
  title={Heat flow in chains driven by thermal noise},
  author={Fogedby, Hans C and Imparato, Alberto},
  journal={Journal of Statistical Mechanics: Theory and Experiment},
  volume={2012},
  number={04},
  pages={P04005},
  year={2012},
  publisher={IOP Publishing}
}

@book{gardiner2009stochastic,
  author={C. W. Gardiner},
  title={Stochastic Methods: A Handbook for the Natural and Social Sciences},
  edition={4},
  publisher={Springer},
  year={2009}
}

@book{risken1996fokker,
  author={H. Risken},
  title={The Fokker--Planck Equation: Methods of Solution and Applications},
  edition={2},
  publisher={Springer},
  year={1996}
}

@article{seifert2012stochastic,
  author={U. Seifert},
  title={Stochastic Thermodynamics, Fluctuation Theorems and Molecular Machines},
  journal={Reports on Progress in Physics},
  volume={75},
  pages={126001},
  year={2012}
}

@book{doi1986theory,
  author={M. Doi and S. F. Edwards},
  title={The Theory of Polymer Dynamics},
  publisher={Oxford University Press},
  year={1986}
}

@article{Esposito2010,
  author={Massimiliano Esposito and Christian Van den Broeck},
  title={Three Detailed Fluctuation Theorems},
  journal={Phys. Rev. Lett.},
  volume={104},
  pages={090601},
  year={2010}
}

@unpublished{Thomas2026Kinetic,
  author = {Jetin E. Thomas and Ramandeep S. Johal},
  title = {Kinetic temperatures and inertial effects in a nonequilibrium bead-spring model},
  note = {arXiv:2608.30809},
  year = {2026}
}

@article{Korn2009,
  author  = {Christian B. Korn and Stefan Klumpp and Reinhard Lipowsky and Ulrich S. Schwarz},
  title   = {Stochastic Simulations of Cargo Transport by Processive Molecular Motors},
  journal = {The Journal of Chemical Physics},
  volume  = {131},
  number  = {24},
  pages   = {245107},
  year    = {2009},
  doi     = {10.1063/1.3279305}
}

@article{Erickson2011,
  author  = {Robert P. Erickson and Zhiyuan Jia and Steven P. Gross and Clare C. Yu},
  title   = {How Molecular Motors Are Arranged on a Cargo Is Important for Vesicular Transport},
  journal = {PLoS Computational Biology},
  volume  = {7},
  number  = {5},
  pages   = {e1002032},
  year    = {2011},
  doi     = {10.1371/journal.pcbi.1002032}
}

@article{Goychuk2014,
  author  = {Igor Goychuk and Vasyl O. Kharchenko and Ralf Metzler},
  title   = {Molecular Motors Pulling Cargos in the Viscoelastic Cytosol: How Power Strokes Beat Subdiffusion},
  journal = {Physical Chemistry Chemical Physics},
  volume  = {16},
  number  = {31},
  pages   = {16524--16535},
  year    = {2014},
  doi     = {10.1039/C4CP01234H}
}

@book{Risken1989,
 author={H. Risken},
 title={The Fokker--Planck Equation},
 publisher={Springer},
 year={1989}
}

@book{Gardiner2009,
 author={C. W. Gardiner},
 title={Stochastic Methods},
 edition={4},
 publisher={Springer},
 year={2009}
}

@book{Bishop2006,
 author={C. M. Bishop},
 title={Pattern Recognition and Machine Learning},
 publisher={Springer},
 year={2006}
}

@misc{Petersen2012,
 author={K. B. Petersen and M. S. Pedersen},
 title={The Matrix Cookbook},
 year={2012},
 note={Version 20121115}
}

@article{Hartich2014,
 author={D. Hartich, A. C. Barato and U. Seifert},
 title={Stochastic Thermodynamics of Bipartite Systems},
 journal={J. Stat. Mech.},
 pages={P02016},
 year={2014}
}

@article{SagawaUeda2010,
  author  = {Sagawa, Takahiro and Ueda, Masahito},
  title   = {Generalized Jarzynski Equality under Nonequilibrium Feedback Control},
  journal = {Physical Review Letters},
  volume  = {104},
  pages   = {090602},
  year    = {2010},
  doi     = {10.1103/PhysRevLett.104.090602}
}

@article{EspositoSchaller2012,
  author  = {Esposito, Massimiliano and Schaller, Gernot},
  title   = {Stochastic Thermodynamics for ``Maxwell Demon'' Feedbacks},
  journal = {Europhysics Letters},
  volume  = {99},
  pages   = {30003},
  year    = {2012},
  doi     = {10.1209/0295-5075/99/30003}
}

@article{MeibohmEsposito2022,
  author  = {Meibohm, Jan Nicolas and Esposito, Massimiliano},
  title   = {Finite-Time Dynamical Phase Transition in Nonequilibrium Relaxation},
  journal = {Physical Review Letters},
  volume  = {128},
  pages   = {110603},
  year    = {2022},
  doi     = {10.1103/PhysRevLett.128.110603}
}

@article{MeibohmEsposito2023,
  author  = {Meibohm, Jan Nicolas and Esposito, Massimiliano},
  title   = {Landau Theory for Finite-Time Dynamical Phase Transitions},
  journal = {New Journal of Physics},
  volume  = {25},
  pages   = {023034},
  year    = {2023},
  doi     = {10.1088/1367-2630/acbc41}
}

@article{Leighton2024,
  author  = {Leighton, Matthew P. and Ehrich, Jannik and Sivak, David A.},
  title   = {Information Arbitrage in Bipartite Heat Engines},
  journal = {Physical Review X},
  volume  = {14},
  pages   = {041038},
  year    = {2024},
  doi     = {10.1103/PhysRevX.14.041038}
}

@article{Szilard1929,
  author  = {Szilard, Leo},
  title   = {{\"U}ber die Entropieverminderung in einem thermodynamischen
             System bei Eingriffen intelligenter Wesen},
  journal = {Zeitschrift f{\"u}r Physik},
  volume  = {53},
  pages   = {840--856},
  year    = {1929},
  doi     = {10.1007/BF01341281}
}

@article{Maruyama2009,
  author  = {Maruyama, Koji and Nori, Franco and Vedral, Vlatko},
  title   = {Colloquium: The Physics of Maxwell's Demon and Information},
  journal = {Reviews of Modern Physics},
  volume  = {81},
  number  = {1},
  pages   = {1--23},
  year    = {2009},
  doi     = {10.1103/RevModPhys.81.1}
}

@article{LeightonSivak2025,
  author  = {Leighton, Matthew P. and Sivak, David A.},
  title   = {Flow of Energy and Information in Molecular Machines},
  journal = {Annual Review of Physical Chemistry},
  volume  = {76},
  pages   = {379--403},
  year    = {2025},
  doi     = {10.1146/annurev-physchem-082423-030023}
}

@article{CaldeiraLeggett1983,
  author  = {Caldeira, A. O. and Leggett, A. J.},
  title   = {Path integral approach to quantum Brownian motion},
  journal = {Physica A},
  volume  = {121},
  number  = {3},
  pages   = {587--616},
  year    = {1983},
  doi     = {10.1016/0378-4371(83)90013-4}
}

@article{DunkelHanggi2009,
  author  = {Dunkel, J{\"o}rn and H{\"a}nggi, Peter},
  title   = {Relativistic Brownian motion},
  journal = {Physics Reports},
  volume  = {471},
  number  = {1},
  pages   = {1--73},
  year    = {2009},
  doi     = {10.1016/j.physrep.2008.12.001}
}

@article{van2004microscopic,
  title={Microscopic analysis of a thermal Brownian motor},
  author={Van den Broeck, Christian and Kawai, Ryoichi and Meurs, Pascal},
  journal={Physical Review Letters},
  volume={93},
  number={9},
  pages={090601},
  year={2004},
  publisher={APS}
}

@article{barato2015thermodynamic,
  title={Thermodynamic uncertainty relation for biomolecular processes},
  author={Barato, Andre C and Seifert, Udo},
  journal={Physical Review Letters},
  volume={114},
  number={15},
  pages={158101},
  year={2015},
  publisher={APS}
}

@article{AllahverdyanJanzingMahler2009,
  author  = {Allahverdyan, Armen E. and Janzing, Dominik and Mahler, G{\"u}nter},
  title   = {Thermodynamic efficiency of information and heat flow},
  journal = {J. Stat. Mech.},
  volume  = {2009},
  number  = {09},
  pages   = {P09011},
  year    = {2009},
  doi     = {10.1088/1742-5468/2009/09/P09011}
}

@article{HerpichShayanfardEsposito2020,
  author  = {Herpich, Tim and Shayanfard, Kamran and Esposito, Massimiliano},
  title   = {Effective thermodynamics of two interacting underdamped Brownian particles},
  journal = {Phys. Rev. E},
  volume  = {101},
  pages   = {022116},
  year    = {2020},
  doi     = {10.1103/PhysRevE.101.022116}
}

@article{TakakiMugnaiThirumalai2022,
  author  = {Takaki, Ryota and Mugnai, Mauro L. and Thirumalai, D.},
  title   = {Information flow, gating, and energetics in dimeric molecular motors},
  journal = {Proc. Natl. Acad. Sci. USA},
  volume  = {119},
  number  = {46},
  pages   = {e2208083119},
  year    = {2022},
  doi     = {10.1073/pnas.2208083119}
}

@article{LeightonEhrichSivak2024,
  author  = {Leighton, Matthew P. and Ehrich, Jannik and Sivak, David A.},
  title   = {Information Arbitrage in Bipartite Heat Engines},
  journal = {Phys. Rev. X},
  volume  = {14},
  pages   = {041038},
  year    = {2024},
  doi     = {10.1103/PhysRevX.14.041038}
}
\end{document}